\documentclass[usenatbib]{mnras}
\usepackage{graphicx}
\usepackage{times,verbatim}
\usepackage[fleqn]{amsmath}
\usepackage{amssymb}
\usepackage{multirow}
\renewcommand{\epsilon}{\varepsilon}
\begin{document}
\title{Kinetic simulations of mildly relativistic shocks I: particle acceleration in high Mach number shocks}
\author[P. Crumley, D. Caprioli, S. Markoff, A. Spitkovsky]{P. Crumley$^{1,2,3}$\thanks{E-mail:
    pcrumley@astro.princeton.edu},
  D. Caprioli$^{4}$, S. Markoff$^{2,3}$, A. Spitkovsky$^1$ \\ 
$^1$Department of Astrophysical Sciences, Princeton University, Princeton, NJ 08554, USA\\
$^{2,3}$Anton Pannekoek Institute for Astronomy, University of Amsterdam \& GRAPPA, P.O. Box 94249, 1090 GE Amsterdam, the Netherlands\\
$^4$Department of Astronomy \& Astrophysics, University of Chicago, Chicago, IL 60637, USA
}

\maketitle
\begin{abstract} 
We use fully kinetic particle-in-cell simulations with unprecedentedly large transverse box sizes to study particle acceleration in weakly-magnetized mildly relativistic shocks traveling at a velocity $\approx 0.75c$ and a Mach number of 15. We examine both subluminal (quasi-parallel) and superluminal (quasi-perpendicular) magnetic field orientations. We find that quasi-parallel shocks are mediated by a filamentary non-resonant (Bell) instability driven by non-thermal ions, producing magnetic fluctuations on scales comparable to the ion gyro-radius. In quasi-parallel shocks, both electrons and ions are accelerated into non-thermal power-laws whose maximum energy grows linearly with time. The upstream heating of electrons is small, and the two species enter the shock front in rough thermal equilibrium. The shock's structure is complex; the current of reflected non-thermal ions evacuates cavities in the upstream which form filaments of amplified magnetic fields once advected downstream.  At late times, 10\% of the shock's energy goes into non-thermal protons and $\gtrsim10\%$ into magnetic fields. We find that properly capturing the magnetic turbulence driven by the non-thermal ions is important for properly measuring the energy fraction of non-thermal electrons, $\epsilon_e$. We find $\epsilon_e\sim 5\times10^{-4}$ for quasi-parallel shocks with $v=0.75c$, slightly larger than what was measured in simulations of non-relativistic shocks. In quasi-perpendicular shocks, no non-thermal power-law develops in ions or electrons. The ion acceleration efficiency in quasi-parallel shocks suggests that astrophysical objects that could host mildly relativistic quasi-parallel shocks -- for example, the jets of active galactic nuclei or microquasars -- may be important sources of cosmic rays and their secondaries, such as gamma-rays and neutrinos.
\end{abstract}
\begin{keywords}
acceleration of particles, plasmas, radiation mechanism: non-thermal, jets
\end{keywords}

\section{Introduction}
\label{sec:Intro}
Collisionless shocks are capable of producing non-thermal particles in power-laws extending for many orders of magnitude in energy. \citet{1977DoSSR.234.1306K},  \citet{1977ICRC...11..132A}, \citet{1978MNRAS.182..147B}, and \citet{1978ApJ...221L..29B} realized that collisionless shocks would be a manifestation of a first-order Fermi/diffusive shock acceleration (DSA) process \citep{1949PhRv...75.1169F}; particles can gain energy from the velocity difference between the upstream and downstream. The particles can reach large energies by being scattered back towards the shock front by magnetic turbulence, thereby crossing the shock front many times. The observed non-thermal radiation from astrophysical systems with strong shocks such as supernova remnants and gamma-ray burst afterglows is strong evidence that shocks in astrophysical systems do indeed accelerate particles.

Classic diffusive shock acceleration theory has been extremely successful at explaining how non-thermal particles can be accelerated into a power-law extending for decades in energy, but it is unable to determine why a particle becomes non-thermal in the first place \citep[see e.g.,][ for reviews]{1983RPPh...46..973D,1987PhR...154....1B, 2001RPPh...64..429M}. That is, a particle may need to be pre-energized before it is  ``injected" into DSA. Because DSA theory is unable to answer how particles are injected, it cannot predict the fraction of the shock's energy that is placed into non-thermal particles. Constraining the non-thermal energy fraction of ions ($\epsilon_p$), electrons ($\epsilon_e$) or magnetic field ($\epsilon_B$) requires either modeling multi-wavelength observations or numerical simulations. While the values of $\epsilon_p,\ \epsilon_e,$ and $\epsilon_B$ depend on several shock parameters (e.g., sonic and Alfv{\'e}nic Mach numbers, magnetic field inclination), both observation and simulations show a striking difference in $\epsilon_e$ between non-relativistic and relativistic shocks. Relativistic shocks appear to create non-thermal electrons more effectively than non-relativistic ones. 

Observational evidence for a changing $\epsilon_e$ with shock velocity comes from a wide range of astrophysical systems. The values obtained are model-dependent, but there is a global trend of increasing $\epsilon_e$ with shock velocity. The non-relativistic shocks of the young supernova remnant Tycho is an inefficient electron accelerator, with $\epsilon_e\sim10^{-3}\epsilon_p$, or $\epsilon_e\sim10^{-4}$ if $\epsilon_p = 0.1$ \citep{2012A&A...538A..81M}.  A low value of $\epsilon_e/\epsilon_p$ in non-relativistic shocks is supported by observations of cosmic rays at Earth. For every cosmic ray electron with $\sim 30\ {\rm GeV}$, several hundred protons are detected \citep[e.g.,][]{PhysRevLett.114.171103,PhysRevLett.113.221102}. On the other hand, the ultra-relativistic shocks that produce gamma-ray burst afterglows are well fit with the much larger $\epsilon_e\sim 0.1$ \citep{1999ApJ...523..177W, 2000ApJ...543...66P, 2014ApJ...785...29S}. Particle-in-cell simulations of electron-ion shocks agree with this trend. The simulations of 1D non-relativistic shocks traveling at $v=0.1c$ in \citet{2015PhRvL.114h5003P} show $\epsilon_e$ of $\sim10^{-4}$ and the relativistic electron-ion shocks in \citet{2011ApJ...726...75S} have an $\epsilon_e\sim 0.1$ at $\gamma=15$.

The production of non-thermal particles depends on the magnetic turbulence at the shock front. In collisionless shocks, binary collisions between particles are not capable of isotropizing the upstream bulk motion of the plasma to subsonic speeds. Instead, the shock must be mediated by electric and magnetic fields that are generated by collective motions in the plasma. There are different mediation mechanisms for shocks, but we focus on the two instabilities that mediate mildly relativistic high Alfv{\'e}nic Mach number shocks: the Weibel instability and the Bell instability. The Weibel instability occurs when initially current-neutral counter-streaming plasma undergoes transverse filamentation into skin-depth wide oppositely directed current filaments, which enhance the transverse magnetic field and eventually isotropize the particles  
\citep{1959PhRvL...2...83W}.
The Weibel instability mediates weakly-magnetized, quasi-parallel, collisionless relativistic shocks \citep{1999ApJ...526..697M, 2008ApJ...673L..39S,2011ApJ...726...75S}. The Bell instability is caused by a current of high-energy ions that are freely streaming through a magnetized plasma. The current in this beam travels along $B_0$ and amplifies transverse waves. The current is balanced by a returning current carried by the background plasma. Because of the different rigidity of energetic ions and compensating electrons, regions experiencing transverse magnetic fluctuations $\delta \mathbf{B}$ lose current balance, and the background plasma feels a $\mathbf{J}\times \delta \mathbf{B}$ force that pushes it outward, growing the fluctuation \citep{2004MNRAS.353..550B,2005MNRAS.358..181B}. 
In its non-linear stage, the Bell instability is capable of producing magnetic fluctuations the size of the  ion's gyroradius, and therefore, it may play an important role in scattering the highest energy particles \citep{2014ApJ...794...46C,2014ApJ...794...47C}.

Mildly relativistic shocks are a good place to accelerate particles to high energies because both the shock speed ($\beta_s$) and Lorentz factor ($\gamma_s$) are of order unity. A large $\beta_s$ means the particle gains more energy with each crossing of the shock-front; $\Delta E/E \propto \beta_s$ in non-relativistic shocks, and $\Delta E/E \sim 1$ after the first crossing in relativistic shocks \citep{1978MNRAS.182..147B, 1989Natur.342..654Q, 2001MNRAS.328..393A}. However, in ultra-relativistic shocks, the component of the magnetic field perpendicular to the shock front is magnified by a factor $\gamma_s$ due to a Lorentz transformation. Therefore, nearly all magnetized ultra-relativistic shocks will be quasi-perpendicular. The quasi-perpendicular magnetic field geometry may severely suppress particle injection in magnetized relativistic shocks because it is superluminal, i.e., particles confined to background magnetic field lines would need to be traveling faster than the speed of light to return upstream \citep{1990ApJ...353...66B,2009ApJ...698.1523S, 2013ApJ...771...54S}. Mildly relativistic shocks with $\gamma_s\beta_s\sim 1$ have an advantage in producing high energy particles because the acceleration time is short, but the shock is subluminal for a larger range of upstream magnetic field inclinations than ultra-relativistic shocks.

To accelerate the highest energy particles, it is not enough to have the particle gain $\Delta E/E\sim 1$ at the shock front. The particles also must return to the shock front quickly enough so they cross the shock many times in a short timescale. The time it takes to accelerate a particle to energy $E$ in a non-relativistic shock depends on the diffusion coefficient $D$ and is $t_{\rm acc} = c^2D(E)/\beta_s^2$ \citep[e.g.,][]{1983RPPh...46..973D}. If the turbulence has large magnetic field fluctuations at all scales, the diffusion coefficient depends linearly on energy, and the maximum energy grows linearly with time. Previous work has shown that when the magnetic turbulence is at small scales, such as the upstream fluctuations that occur in unmagnetized shocks, the maximum energy increases more slowly as $\sqrt{t}$ \citep[e.g.,][]{2013ApJ...771...54S}.  The difference between the two scalings may be the difference in whether an object can be an ultra-high energy cosmic ray progenitor. 

This paper is the first in a series of papers that kinetically examines mildly relativistic shocks ($\gamma_s\beta_s\sim 1$). In this paper we show the necessary ingredients to properly capture the magnetic field structure at the shock front in a 2D simulation of weakly magnetized, electron-ion shocks and to measure $\epsilon_e$. We examine mildly relativistic shocks traveling at $\gamma\beta\approx 1$ using self-consistent, fully kinetic, particle-in-cell simulations. Most previous studies of the mildly relativistic shocks have used Monte-Carlo and semi-analytical methods which require assumptions about the turbulence and the fraction of non-thermal particles \citep{2000ApJ...542..235K, 2005PhRvL..94k1102K, 2007ApJ...658.1069M, 2012ApJ...745...63S, 2013ApJ...776...46E}, or fully kinetic simulations that were too short to study the late-time evolution of the shock structure \citep{2006NJPh....8..225D, 2017ApJ...847...71B,  2018MNRAS.473..198D}. We do not vary the shock velocity in this paper, but we plan on using the results of this paper to examine how $\epsilon_e$, $\epsilon_B$, and $\epsilon_p$ change in the transition between non-relativistic and relativistic shocks in a future work.

Our paper is organized as follows: first, we describe our simulation setup in Section \ref{sec:Setup}. In Sections \ref{sec:FiducialRun} and \ref{sec:ParticlAcc}, we analyze the structure of and particle acceleration in a mildly relativistic quasi-parallel shock. In Section \ref{sec:WideEnough}, we show how resolving the ion gyro-radius is important to capture the turbulence at the shock front and measuring $\epsilon_e$. In Section \ref{sec:SuperLum}, we show that superluminal shocks are unable to accelerate particles into a power-law. Finally we conclude by summarizing our results and discussing their astrophysical relevance. Many astrophysical objects  host mildly relativistic outflows capable of producing strong shocks. However, the magnetized mildly relativistic shocks studied in this paper are most relevant to the internal shocks in jets with modest Lorentz factors like the jets observed as bright sources of non-thermal radiation in microquasars and low luminosity active galactic nuclei (AGN).
	
\begin{figure}
\includegraphics[width=0.45\textwidth]{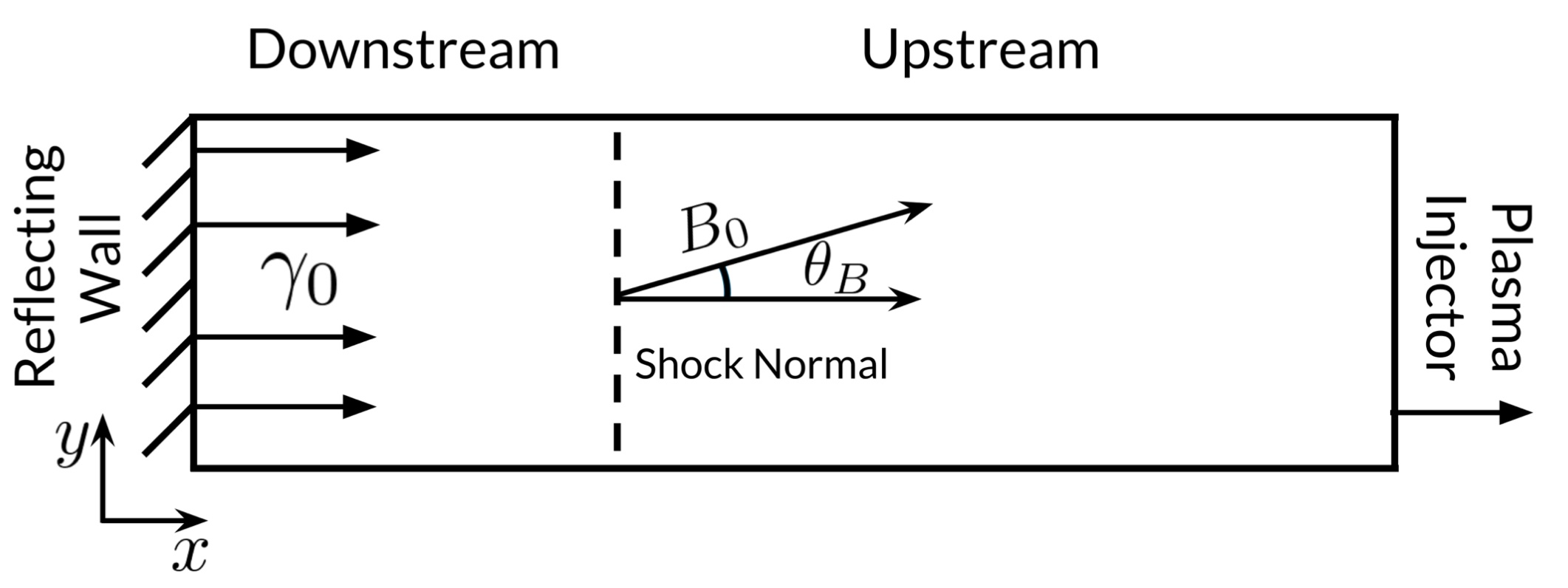}
\caption{A schematic of our shock simulation. Plasma is injected at rest at the right boundary and a reflecting wall is pushed into the plasma with Lorentz factor $\gamma_0$, forming the shock. The injected plasma is initialized with a background magnetic field that lies in the $x-y$ plane. We use periodic boundaries in the $y$ direction}
\label{fig:Setup}
\end{figure} 

\section{Methodology}
\label{sec:Setup}

We use the parallel electromagnetic PIC code \textit{TRISTAN-MP} \citep{OriginalTristan, 2005AIPC..801..345S} to simulate mildly relativistic electron-ion shocks \citep{OriginalTristan, 2005AIPC..801..345S}. The computational domain is a 2D Cartesian $xy$ grid, with length $L_x$ and width $L_y$. While the computational domain is 2D, we track all three spatial components of the field quantities and the particle momenta. We use periodic boundary conditions along the $y$ direction and we place reflecting walls at the left and the right boundary of the computational domain. A schematic of our simulation setup is shown in Figure \ref{fig:Setup}. 

Our simulation setup is very similar to previous PIC simulations of collisionless shocks, e.g., \cite{2008ApJ...673L..39S, 2011ApJ...726...75S, 2013ApJ...771...54S},  with one major difference: the simulation's restframe. Our simulations are performed in the upstream rest frame while previous simulations were performed in the downstream restframe. We find that having a stationary upstream helps reduce numerical heating that occurs when simulating a  moving cold plasma beam. At the right boundary, we inject electrons and ions drawn from stationary Maxwell-J{\"u}ttner distributions with thermal spread $\Delta\gamma \equiv  m_i c^2/k_B T$, where $k_B$ is the  Boltzmann constant, $T$ is the temperature, and $m_i$ is the ion mass. The two species are in thermal equilibrium with $T_i=T_e$. The left wall travels to the right with a Lorentz factor $\gamma_0 = 1.5$, $v_0\approx 0.75c$, driving a shock. We truncate the computational domain left of the wall, and measure all distances relative to the left wall.\footnote{Measuring the distances relative to the moving left wall causes the upstream to appear to be moving to the left with time even though its velocity is zero in the simulations.} The right boundary expands and we do not let any shock reflected particles reach the right boundary. To compare our results to downstream simulations, we Lorentz boost some quantities like the downstream spectra to the left-wall restframe.

{\it TRISTAN-MP} uses a normalized unit system, like most particle-in-cell codes. Time is measured in units of the upstream inverse plasma frequency, $\omega_{pe}^{-1} = \sqrt{m_e/(4\pi n q^2)}$, where $m_e$ is the electron mass, $n$ is the upstream number density of electrons, and $q$ is the charge of electrons. Distances are measured in units of the electron skin depth in the upstream plasma, $c/\omega_{pe}$. We report lengths in units of the ion skin depth,  $c/\omega_{pi} = c\omega_{pe}^{-1}\sqrt{m_i/m_e}$, and we measure times in terms of the upstream ion plasma frequency ($\omega_{pi}^{-1}$). We use a reduced mass ratio, $m_i/m_e$, set to 64 in the fiducial run, easing computational expense of reaching late times in terms of $\omega_{pi}t$. We checked for convergence in mass ratio and show our results in Appendix \ref{sec:MassRatio}. We resolve the upstream electron skin depth with a different number of cells depending on the simulation. We use the criterion that we must resolve both the upstream Debye length and the downstream electron skin depth with at least one cell. The Debye length is $\lambda_D = c\omega_{pe}^{-1}\sqrt{\frac{m_i}{m_e} \Delta \gamma}$.  In the fiducial run, this criterion is satisfied when resolving the upstream electron skin depth with 4 cells, with a timestep of $\Delta t=0.1125 \omega^{-1}_{pe}$. The low resolution is required to capture the relevant ion length and timescales. We use 4 particles per cell (2 ions and 2 electrons) and filter particles' contributions to the current 32 times per timestep to reduce noise, as is routinely done \citep[e.g.][]{2005AIPC..801..345S}. We checked convergence in the number of particles-per-cell (ppc) and spatial resolution, running smaller simulations with ppc = 64 and $c/\omega_{pe}=8$.

We initialize an upstream magnetic field $B_0$ with inclination $\theta_B$ to the shock normal. A completely parallel shock has $\theta_B = 0$ and a completely perpendicular shock has $\theta_B=90^\circ$. We use two different magnetic field orientations in this paper, one is quasi-parallel ($\theta_B\approx 10^\circ$), and the other is quasi-perpendicular ($\theta_B \approx 55.^\circ$). In the downstream rest frame the angle, $\theta'_B$, is $15^\circ$ and $65^\circ$ respectively (\(\gamma_0\tan{\theta_B}=\tan{\theta'_B}\)). 
We choose a magnetic field orientation such that perpendicular component of the initialized magnetic field lies entirely in the $xy$-plane, i.e., $B_{z,0} = 0$. 

The strength of the magnetic field is described by $\sigma_0$, the ratio of the energy density of the upstream magnetic field to the kinetic energy of the upstream flow as measured in the downstream frame: $\sigma_0 = v'^2_\mathcal{A}/(c^2[\gamma_0-1])$ where $v'_\mathcal{A}$ is the upstream Alfv{\'e}n velocity as measured in the downstream frame. Re-writing everything in upstream quantities:
\begin{equation}
\label{eq:Sigma}
\sigma_0 =\frac{B_0^2 (\gamma_0^2 \sin^2{\theta_B}+\cos^2{\theta_B})}{4\pi \gamma_0 n_0 m_i c^2(\gamma_0-1)(1+m_e/m_i)},
\end{equation}
In this paper, we fix $\sigma_0 = 0.007$.  The magnetization fixes the Alf{\'e}nic Mach number, which is the shock speed divided by upstream Alfv{\'e}nic velocity. Using conservation of particle number $\beta_s = r\beta_0/(r-1)$, where $r$ is the compression ratio of the shock, so the Alfv{\'e}nic Mach number is
\begin{equation}
M_\mathcal{A}\approx\frac{r}{r-1}\frac{\beta_0}{\sqrt{\sigma_0(\gamma_0-1)}}
\end{equation}
Since our shock is strong, the compression ratio of the shock, $r\sim 4\gamma_0^2$ in the upstream frame and for $\beta_0 = 0.75$, $\sigma_0=0.007$, we find $\beta_s \approx 0.83$ and $M_\mathcal{A}\approx 15$. 

Plasma beta, $\beta_p$, is the ratio between the thermal pressure and the magnetic pressure in the upstream plasma:
\begin{equation}
\beta_p  = \frac{4c^2\Delta\gamma}{v^2_\mathcal{A}\left(1+\frac{m_e}{m_i}\right)},
\end{equation}
where $v_\mathcal{A}$ is the Alfv{\'e}nic velocity in the upstream frame. Increasing the plasma beta for a fixed magnetization will decrease the sonic Mach number of the shock. We assume equilibrium between the magnetic pressure and thermal pressure of the upstream plasma, that is $\beta_p = 1$. For $\gamma_0=1.5$, $M_\mathcal{A}=15$, $m_i/m_e=64$, and $\theta'_B =15^\circ$,  $\beta_p =1$ corresponds to a temperature $\Delta\gamma = 1.27\times10^{-3}$ and sonic Mach number $M_s\approx15$. 

We measure three quantities that capture how much of the upstream kinetic energy goes into magnetic fields, non-thermal ions, and non-thermal electrons: $\epsilon_B,\ \epsilon_p,$ and $\epsilon_e$, respectively. To calculate $\epsilon_B$, we Lorentz boost the total magnetic field $\bf B$ into a frame moving at $\gamma_0$ in the $+x$ direction and use
\begin{equation}
\label{eq:epsB}
\epsilon_B =  \frac{|{\bf B}'|^2(1+m_e/m_i)^{-1}}{4\pi \gamma_0 n_0 m_i c^2(\gamma_0-1)}.
\end{equation}
$\epsilon_B$ and $\sigma_0$ are the only field quantities we report in the downstream rest frame. We measure $\epsilon_p$ and $\epsilon_e$ by first Lorentz boosting the spectra into the downstream rest frame and then measuring the fraction of energy in the non-thermal particles. This ratio is converted to $\epsilon_p$ and $\epsilon_e$ by multiplying by the average energy carried by the respective particle downstream and dividing by the average kinetic energy of an incoming ion. Or more explicitly:
\begin{equation}
\epsilon_{p,e} = \frac{\int_{E_{\rm inj}}^{\infty}{E'\frac{\mathrm{d}n'}{\mathrm{d}E'} dE'}}{\int_0^{\infty}{ E'\frac{\mathrm{d}n'}{\mathrm{d}E'} dE'}}\times \frac{m_{i,e} (\langle\ \gamma_{i,e}'\rangle-1)}{m_i (\gamma_0-1)}.
\end{equation}

The transverse size of our fiducial run is 16000 cells, equivalent to 4000 electron skin depths or 500 ion skin depths for the reduced mass ratio $64$, significantly wider than previous PIC shock simulations \citep[e.g.,][]{2009ApJ...698.1523S,2011ApJ...726...75S,2012ApJ...759...73N,2013ApJ...771...54S,2015PhRvL.114h5003P}. For example, while \citet{2017ApJ...847...71B} also studied electron acceleration in high Mach number, mildly relativistic shocks, our simulations are $\sim20$ times wider in terms of ion skin depths and were run 3 times longer in terms of ion gyrofrequency (even when accounting for the different rest frames).
The size of our simulations are more comparable to hybrid-MHD (kinetic ions-fluid electrons) simulations. We require a large size to resolve the gyroradii of the non-thermal ions. We summarize the simulations run in this work
in Table \ref{tab:Summary}.
\begin{table}
\caption{Summary of Simulations. See Section \ref{sec:Setup} for more information.}
\label{tab:Summary}
\begin{tabular}{l cccc}
\hline
%$B_0$ Inclination & $B_0$ Orientation & $m_i/m_e$ & $L_y/\Delta$ & $c/\omega_{pe}/\Delta$ \\
$B_0$ Inclination & $B_0$ Orientation & $m_i/m_e$ & $L_y$ & $c/\omega_{pe}$ \\
 & & & \# of cells & \# of cells \\
\hline
\multirow{5}{1.5cm}{Quasi-para $10.13^\circ$}
 & in-plane & 64 & 16000 & 4\\
 & in-plane & 64 & 800 & 4\\
 & in-plane & 16 & 3500 & 8\\
 & in-plane& 160 & 4000 & 4\\
 & out-of-plane & 64 & 3220 &  4\\
\hline
\multirow{2}{1.5cm}{Quasi-perp $55^\circ$}
 & in-plane & 64 & 1600 & 4\\
 & out-of-plane &  64 & 1600 & 4\\

\hline
\multicolumn{5}{c}{All simulations had $\sigma = 0.007$ and $\beta_p=1$ ($M_A\approx M_s\approx 15$).}
\end{tabular}

\end{table}
\section{Structure of a Bell-mediated shock.}

\label{sec:FiducialRun}
\begin{figure}
\centering
\includegraphics[width=0.5\textwidth]{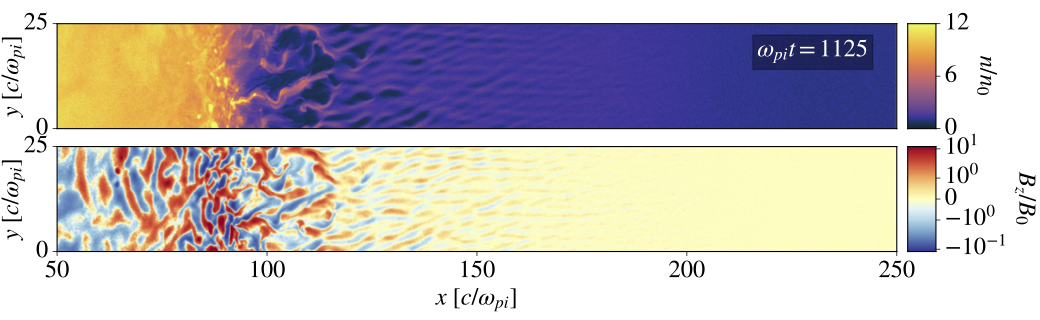}
\includegraphics[width=0.5\textwidth]{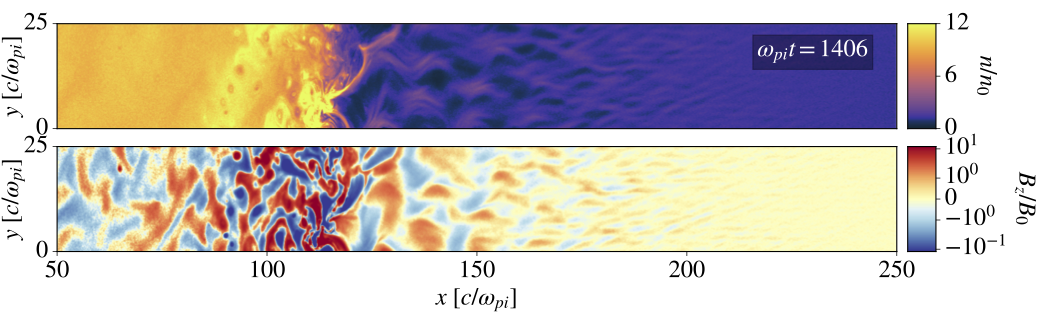}
\includegraphics[width=0.5\textwidth]{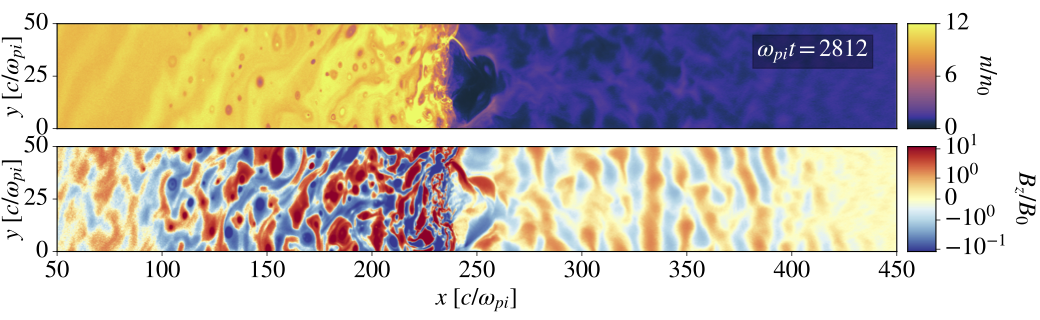}
\includegraphics[width=0.5\textwidth]{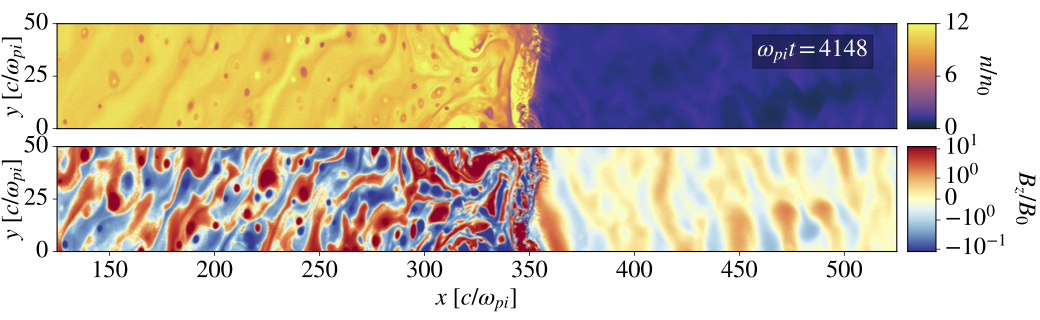}
\caption{The transition from a Weibel mediated shock to Bell-mediated shock. A small section of the simulation around the shock is followed with time. Initially the upstream filaments are on the order of a few ion skin depths with similar sized magnetic perturbations. As the shock becomes Bell-mediated, cavities are evacuated upstream and advected downstream. The magnetic field amplification is significantly larger at the walls of the cavities while the center of the cavities has a small magnetic field. The size of the cavities grows with time. At $\omega_{pi}t=4148$, there are no cavities close to the shock front in this section of the simulation (one is forming around $x \sim 475c/\omega_{pi}$), and the magnetic amplification is large enough such that the shock is locally quasi-perpendicular with a small Alfv{\'e}nic Mach number. Whistler waves can be observed at the shock, see inset. Whistler waves may be an effective preheating source for injection of electrons \citep{2011ApJ...733...63R}.}
\label{fig:Transition}
\end{figure}

\begin{figure}
\centering
\includegraphics[width=0.5\textwidth]{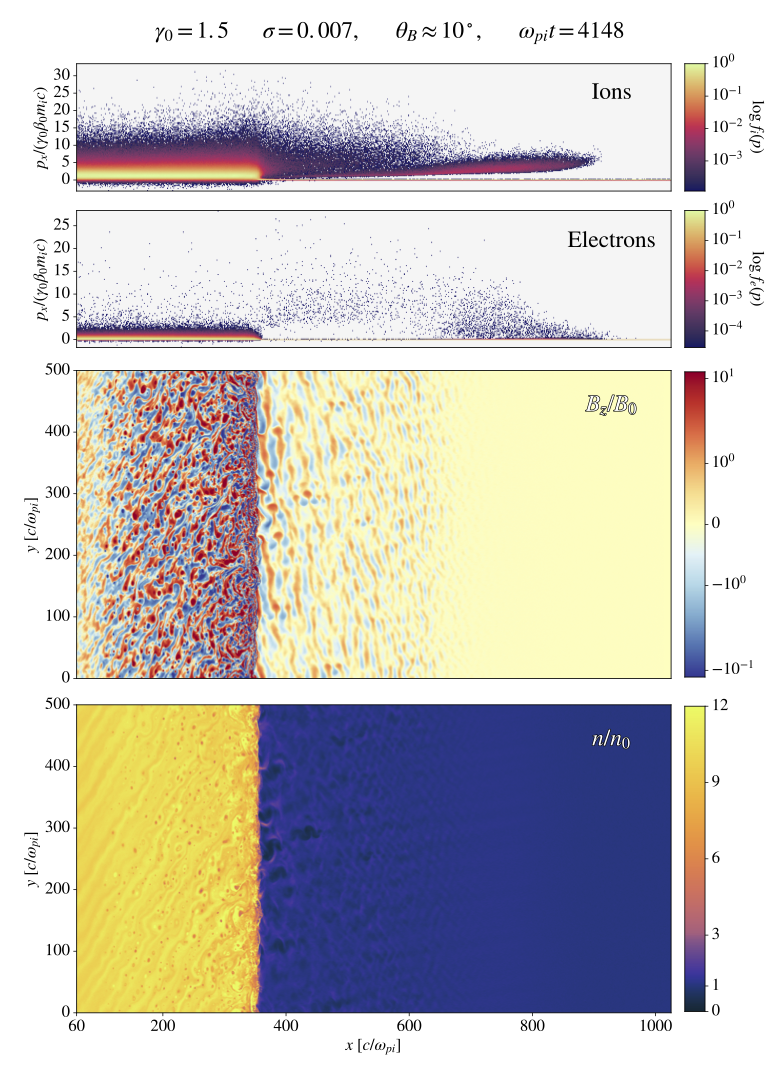}
\caption{The fiducial run at time $\omega_{pi}t=4148$. Distances are measured in ion skin depths. The top two graphs are $x$---$p_x$ phase diagrams for the ions and electrons respectively. The bottom two graphs show $B_z$ and the density domain. The upstream location	 where Bell instability starts to grow ($x\sim650 c/\omega_{pi}$) is visible in all four subplots. In the ion phase diagram it is where the ions transition from a narrow beam in phase space to a more diffuse, scattered structure. In the electron phase diagram, Bell is marked by the dearth of reflected electrons with energies below $\sim \gamma^2_0 m_i c^2$ between $350c/\omega_{pi}$ and $650c/\omega_{pi}$ . These two features are aligned with the transverse, circularly polarized magnetic field amplification seen in the $B_z$ plot, and the corresponding evacuated cavities in the density.
}
\label{fig:BigOne}
\end{figure}
\begin{figure}
\centering
\includegraphics[width=\linewidth]{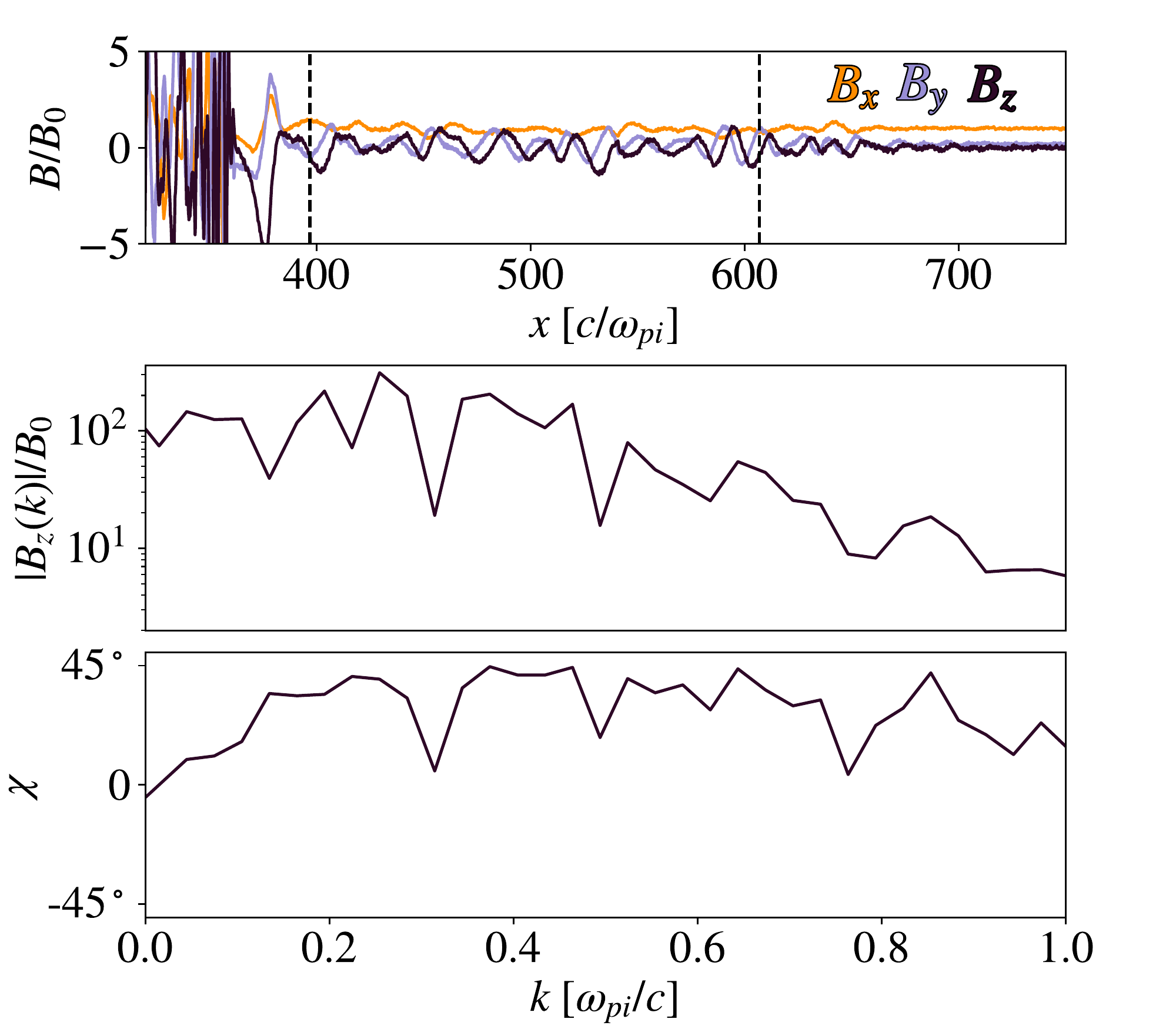}
\caption{A measurement of the polarization of the upstream waves at time $\omega_{pi}t=4148$. The top panel shows a 1D slice of the 3 magnetic field components. Vertical dashed lines mark the region where the Fourier transform of the fields is calculated. The middle panel shows the Fourier transform of $B_z$, and the bottom panel is a graph of Stokes polarization angle $\chi$. Non-resonant Bell waves are right handed circularly polarized ($\chi = 45^\circ$), as are the waves in the upstream of the simulation.}
\label{fig:FFT}
\end{figure}

\begin{figure*}
\centering
\begin{tabular}{@{}l@{}l@{}}
\includegraphics[width=0.5\linewidth]{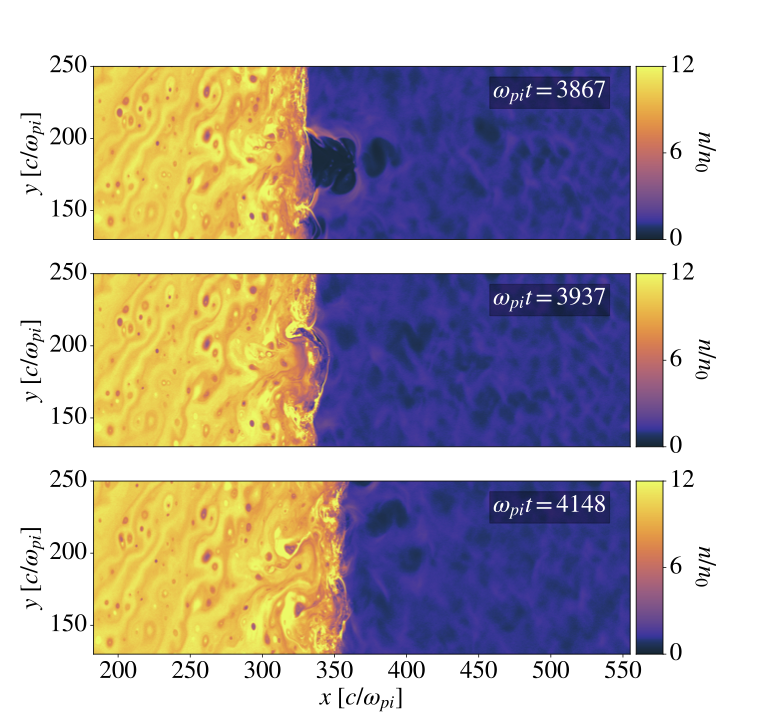}
& \includegraphics[width=0.5\linewidth]{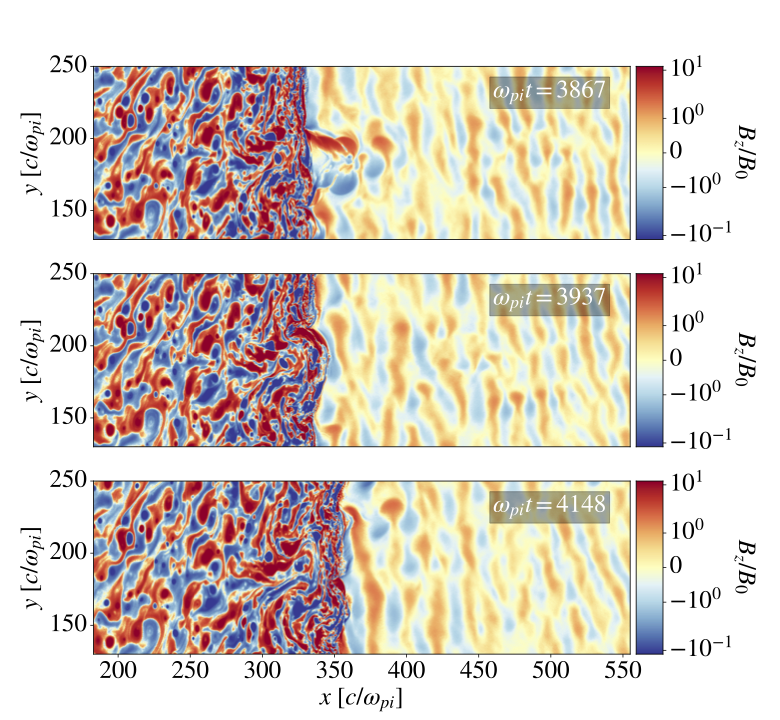}%\\
\end{tabular}
  \caption{Advection of an evacuated upstream cavity into downstream. The left panels show the density and the right show the out-of-plane magnetic field $B_z$. Once advected downstream, the upstream cavities forms two holes with opposite $B_z$ sign filled with a large amount of magnetic field.}
\label{fig:Filaments}
\end{figure*}

\begin{figure}
\centering
\includegraphics[width=\linewidth]{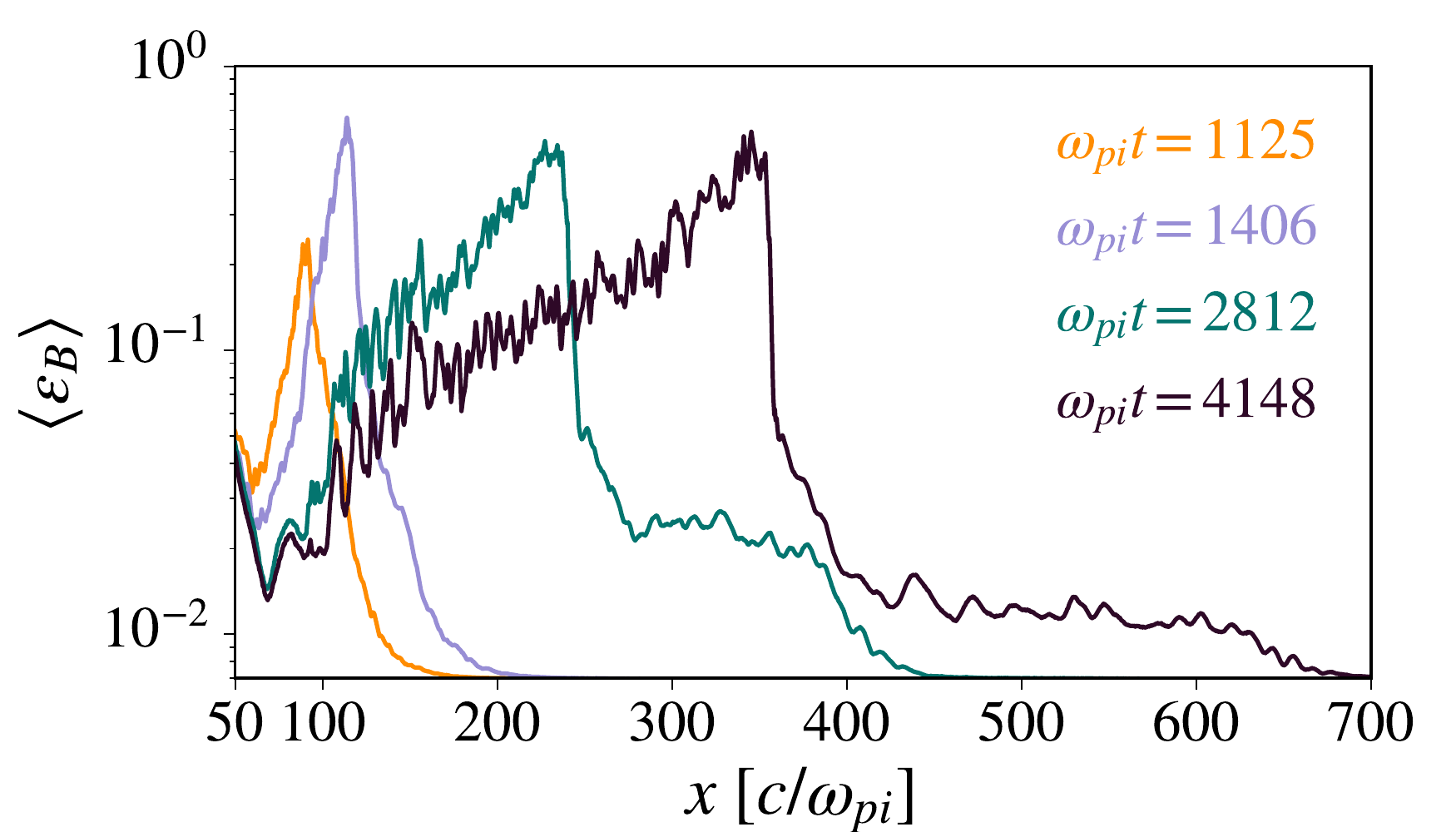}
\caption{The evolution of the y-averaged $\epsilon_B$ with time in the fiducial run. The times correspond to the times shown in Figure \ref{fig:Transition}. After the shock becomes mediated by the Bell instability ($\omega_{pi}t\sim1500$), the magnetic field amplification at the shock front increases. In addition, the amplified magnetic field increases further upstream and downstream with time.  Note: the increase of $\epsilon_B$ at $x\lesssim 60\ \omega_{pi}$ is an artifact from the initialization of the shock}
\label{fig:epsBvt}
\end{figure}
In this section we discuss the structure of a quasi-parallel, electron-ion, magnetized, mildly-relativistic shock. With our large simulations, we are able to capture the large-scale structure of magnetic turbulence upstream. For the first time in a PIC simulation, we use a large enough box to resolve the gyroradius of the highest energy ion as the shock transitions between the early times---when the Weibel instability mediates the shock---to late times---when Bell instability is the dominant instability. 

The evolution of density and $B_z$ with time is shown in Figure \ref{fig:Transition} for a small slice of our simulation. The shock forms because of the Weibel instability, which starts the Fermi process. Once they have escaped upstream, particles are scattered back to the shock front due to small angle scattering, accelerating protons and electrons \citep{2008ApJ...682L...5S,2011ApJ...726...75S}. Sufficiently high-energy particles are not efficiently scattered by Weibel instability, and are able to escape far upstream. There is a net positive current $\mathbf{J}$ carried by non-thermal ions that non-resonantly grows transverse $\mathbf{B}$ waves, and the shock transitions from being Weibel-mediated to Bell-mediated. 

The growth rate of the fastest growing wavelength of the Bell instability depends only on the current carried by non-thermal ions upstream, which are traveling at roughly the shock speed $\beta_s$. In the upstream frame the growth time of the Bell waves is given by eq (16) of \citet{2006PPCF...48.1741R}:
\begin{equation}
\label{eq:BellGrowthTime}
\omega_{pi}t_{\rm Bell} \approx \frac{2}{\xi_{cr}\beta_s}
\end{equation}
$\xi_{cr}$ is the number fraction of non-thermal ions in the upstream, and $\beta_s \approx 0.83$ is the shock velocity. Using the measured parameters from our simulation, i.e., $\xi_{cr}\sim 0.01$, the growth time is $\sim 60/\omega_{pi}$. transverse waves start to appear around $\omega_{pi}t\sim 560$, and the shock completely transitions to being  Bell-mediated around $\omega_{pi}t\sim 1500$ or $\sim 25$ times the growth time of the Bell instability. The Bell modes are visible as a striped $B_z$ and $B_y$ pattern upstream, transverse to the shock normal (see Figures \ref{fig:Transition} \& \ref{fig:Filaments}). We measured the polarity of the upstream waves and found they are right hand circularly polarized, as expected for non-resonant ion waves (see Figure \ref{fig:FFT}). 

The ion current $\mathbf{J}$ also drives a filamentary mode.  Our simulation shows a similar structure to the one found in hybrid simulation of non-relativistic quasi-parallel shocks \citep[see][]{2013ApJ...765L..20C}. To keep the plasma quasi-neutral, there is a returning negative current in the background plasma that is aligned with $B_0$. Therefore, the background plasma feels a $-\mathbf{J}\times\delta\mathbf{B}$ force, which pushes the background plasma away from the regions of strongest current and focuses the current carrying ions, growing the instability further. The size of these upstream cavities grows with time, and they contain a large magnetic field at their walls. The transverse size is limited to be less than the size of gyro-radius of the current-carrying ions.

We show the entire simulation domain of our fiducial, magnetized quasi-parallel shock at the final time step $\omega_{pi}t=4148$ in Figure \ref{fig:BigOne}. The upper two panels show the ion and electron $x$---$p_x$ phase diagrams, and lower two panels show $B_z$ and the density. The upstream location where the reflected ions start to non-resonantly drive circularly-polarized transverse magnetic field fluctuations, $x\sim650 c/\omega_{pi}$, is also visible in structural changes in the top two panels of the phase diagrams. The reflected ions transition from being a relatively narrow beam of particles to having a more diffuse structure as a result of the efficient scattering by the magnetic field. In the electron phase diagram the transition is marked by a minimum momentum required to participate in the particle acceleration. The minimum momentum is visible as a lack of non-thermal electrons with Lorentz factors $\lesssim \gamma^2_0 m_i/m_e$ between $350c/\omega_{pi}$ and $650 c/\omega_{pi}$. All of the non-thermal electrons have energies comparable to the non-thermal ions. The upstream magnetic field amplification is accompanied by the evacuation of small cavities in the density far upstream that grow to larger worm-like cavities closer to the shock front. The amplified magnetic fields are largest at the walls of these cavities. Downstream, the cavities become long filaments with $\epsilon_B\gtrsim 0.2$ and small hole of low densities and large $B_z$, $\epsilon_B\gtrsim 1$.

 When swept downstream, the cavity forms two underdense holes that contain large magnetic fields of opposite $B_z$ polarity, and are surrounded by a ring of current. A close-up of a cavity being advected downstream can be seen in Figure \ref{fig:Filaments}. In 3D, the two holes would likely be connected by a flux tube. Some holes merge downstream, and in 3D, the disruption of possible flux tubes may be an important site of magnetic dissipation. Whether the flux tubes do indeed form in 3D simulation, and the importance of any magnetic dissipation during their lifetimes is left to future work.
 
At the latest times, whistler waves are observed in the ion foot of the shock. Whistler waves are small-scale, dispersive waves that can occur in quasi-perpendicular shocks with low Alfv{\'e}nic Mach numbers. Whistler waves can grow in non-relativistic quasi-perpendicular shocks when $M_\mathcal{A}\lesssim \sqrt{m_i/m_e}$ \citep[e.g.][]{2002PhPl....9.1192K,2011ApJ...733...63R}. This inequality is satisfied in our simulation when $\delta B_\perp/B_0\gtrsim 2$, which is indeed the case in the regions where whistler waves appear in our simulation (see inset of bottom panels of Figure \ref{fig:Transition}. Whistler waves may play an important role in electron injection, as they are an efficient way to transfer energy from ions to electrons \citep{2011ApJ...733...63R}. 

The change in magnetic field amplification with time is best seen in Figure \ref{fig:epsBvt}. After the shock becomes Bell-mediated, at $t\sim 1500\omega_{pi}$, there is significantly larger magnetic field amplification both upstream and downstream compared to earlier times when the shock was Weibel mediated. The extent of the amplification both upstream and downstream grows with time. The average value of magnetic field amplification saturates at $\sim2B_0$ upstream with a peak value of $\gtrsim 5 B_0$ inside of the filaments close to the shock front. We note that $\epsilon_B$ upstream decreases by $\sim33\%$ from $\omega_{pi}= 2812$ and $\omega_{pi}t= 4148$.  This decrease may be because the initial beam of ions that triggers the Bell instability is more anisotropic than the later reflected ions which are scattered more efficiently. The change from a beam to a diffuse structure in phase space is clearly visible in the ion phase diagram in Figure \ref{fig:BigOne}. At later times, the importance of the organized beam will decrease and the reflected particles may be more diffuse.

\section{Particle acceleration in a Bell-mediated  shock}
\label{sec:ParticlAcc}
\begin{figure}
\centering
  \includegraphics[width=\linewidth]{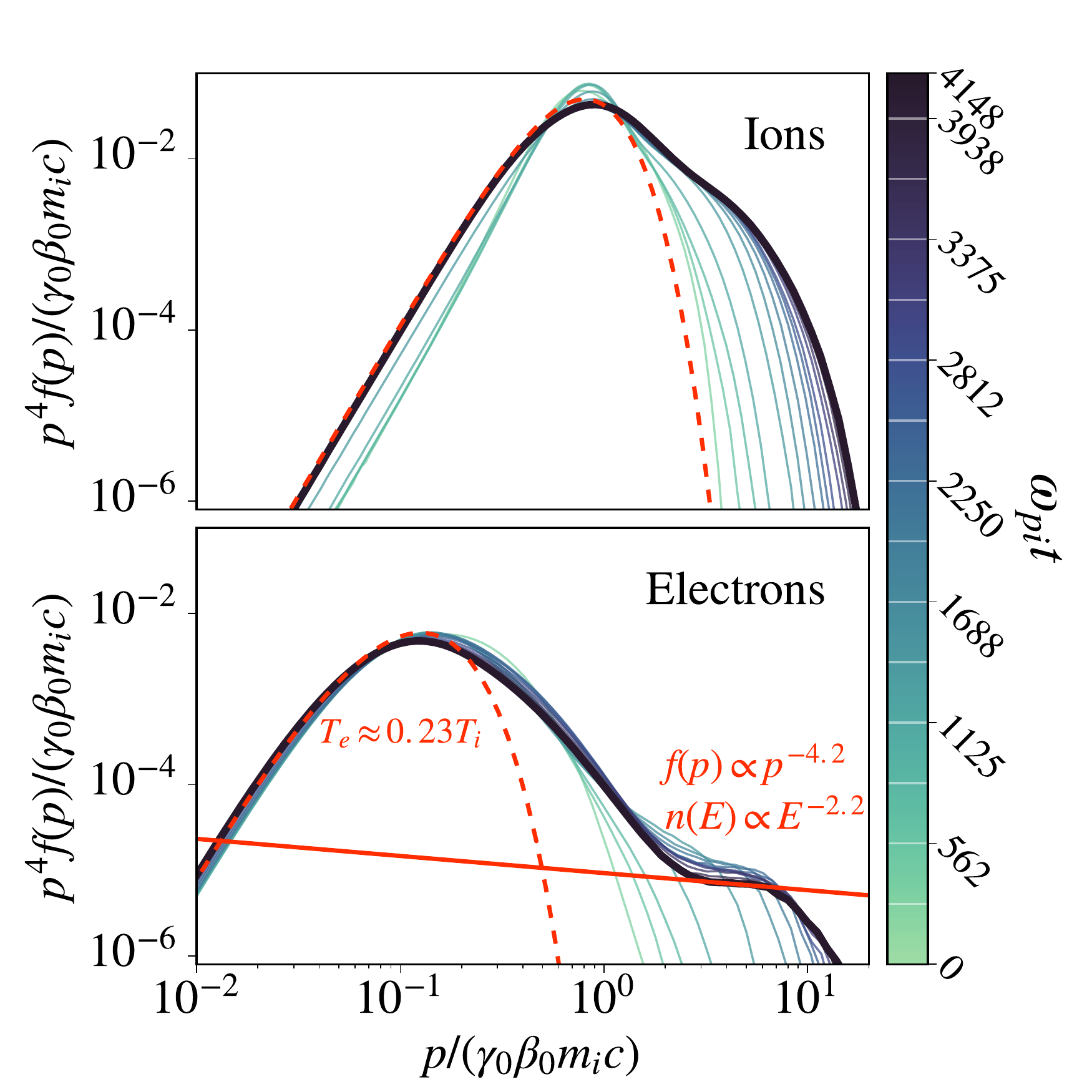}
  \caption{Time evolution of the downstream spectrum for the fiducial run. The spectra are boosted into the downstream rest frame and extracted from an area $10-80c/\omega_{pi}$ downstream of the shock. When the shock had not yet traveled far enough to the right, we take the leftmost section of the downstream, between $\max(0,x_s-200)$ and $\max(5,x_s-100)$. The white lines in the color bar mark the times when spectra are plotted, including the final tick mark. The ions have a temperature $\sim 0.15 m_i c^2$, and the electrons equilibrate at a temperature $\sim23\%$ of the ions.}
  \label{fig:SpectEvo}
\end{figure}

\begin{figure}
  \includegraphics[width=\linewidth]{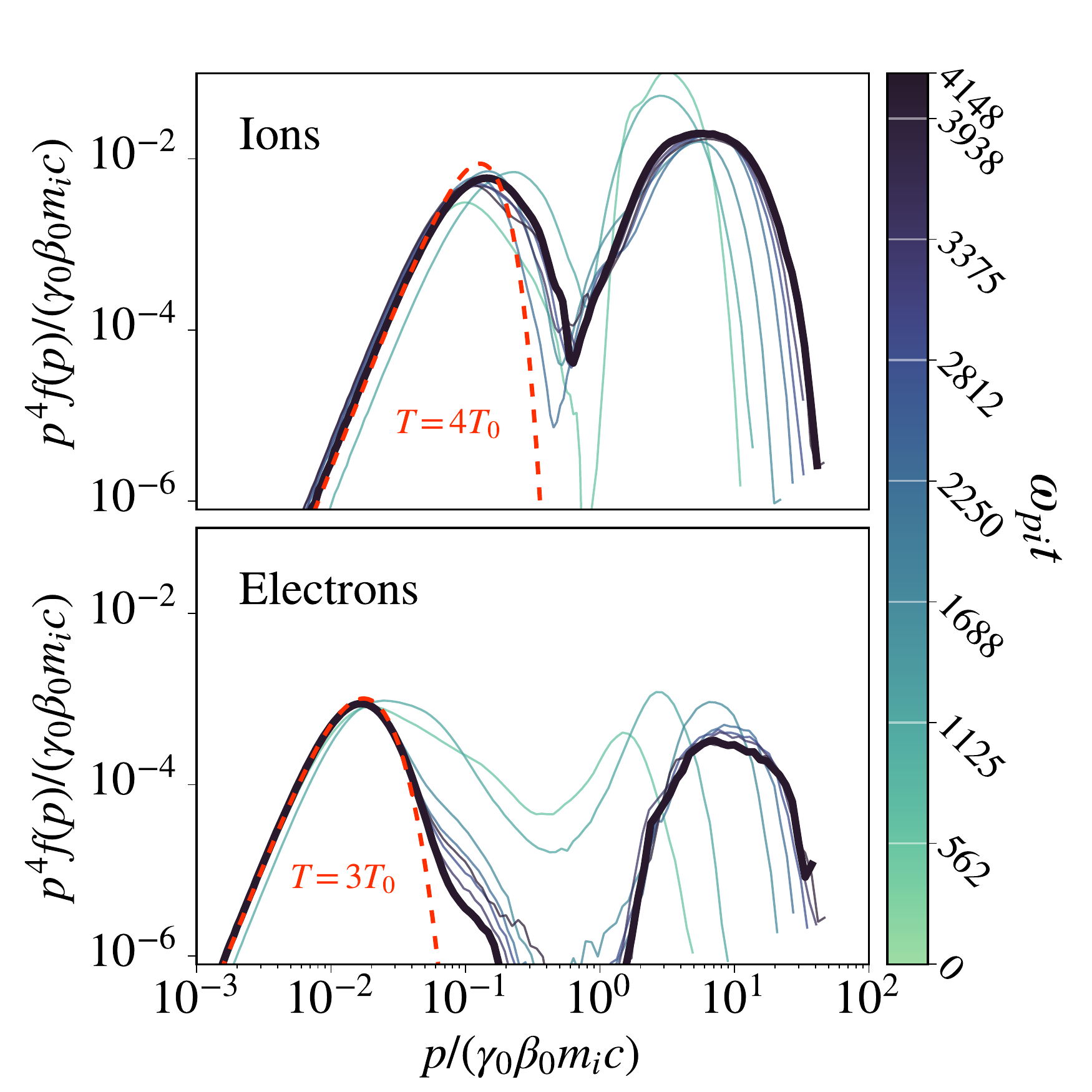}
\caption{Time evolution of the upstream spectrum  for the fiducial run. The spectra shown in the simulation (i.e., upstream) rest frame and extracted from an area $10-20c/\omega_{pi}$ upstream of the shock. Both the ions and electrons show two-humped spectra. The left hump is the incoming beam of particles while the right hump is the reflected particles. In the spectra we show Maxwellian distributions fit-by-eye, with temperatures in terms of $T_0$, the injected temperature of the plasma. When the shock is Weibel mediated, the electrons are preheated significantly, which can be seen because the left hump is not fit well by a Maxwellian at early times. After Bell instability kicks in at around $t \sim 1500 \omega_{pi}$ both the ions and electrons are preheated only by a factor $\sim$ a few.}
  \label{fig:UpSpectEvo}
\end{figure}

\begin{figure}
\centering
\includegraphics[width=\linewidth]{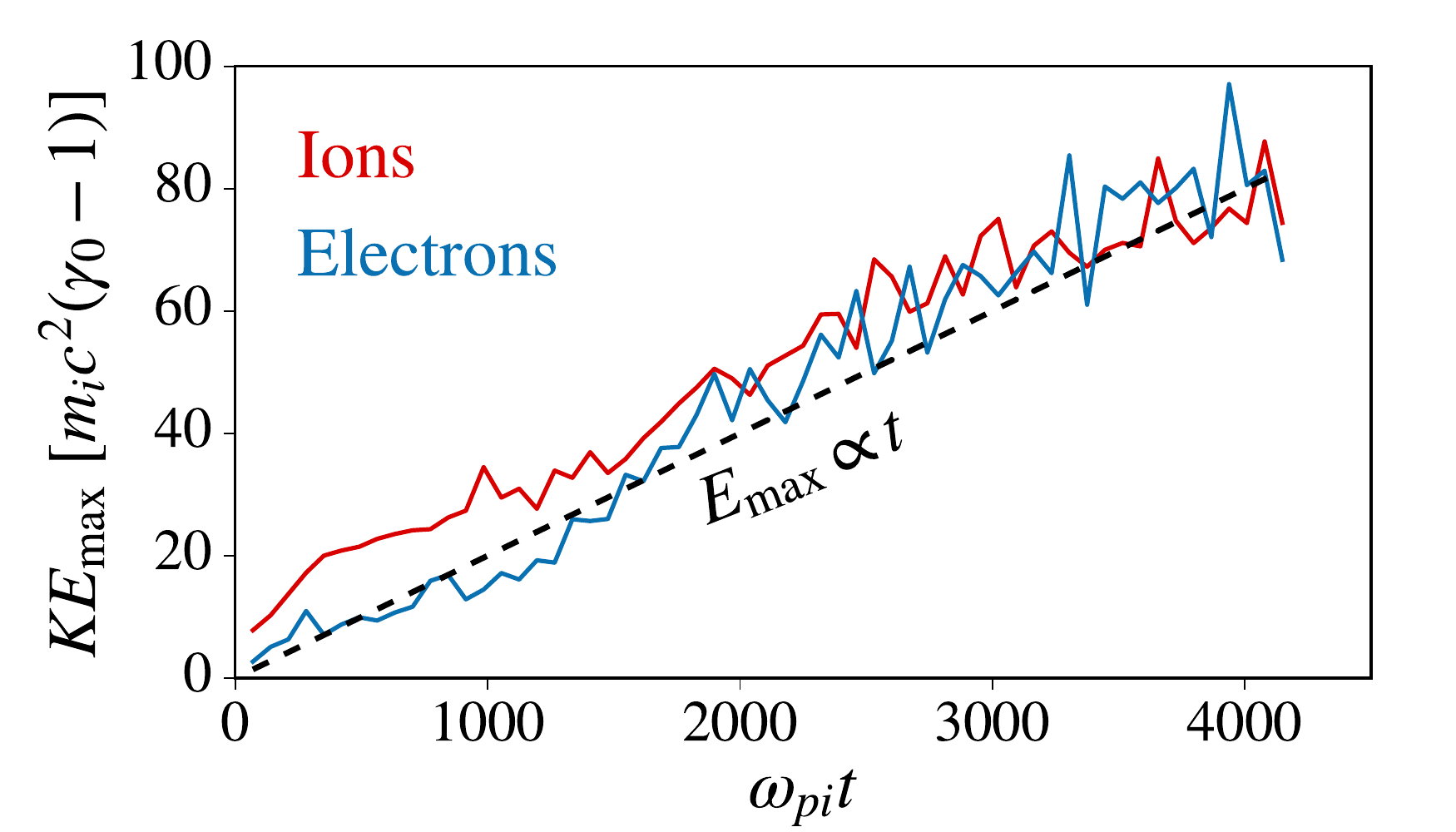}
\caption{The energy of the highest energy ion and electron is plotted as a function of time. After the shock becomes Bell-mediated (around $\omega_{pi}t\sim1500$),  the maximum energy of both the electrons and ions is comparable to each other and increases linearly in time with no sign of saturation, consistent with the Bohm scaling.}
\label{fig:Emax_v_t}
\end{figure}

In the previous section, we showed that the transition from a Weibel-mediated shock to Bell-mediated shock changes the upstream magnetic field turbulence and shock geometry. In this section, we examine the implications of implications of the Bell turbulence on particle acceleration. 

The downstream spectral evolution with time is shown in Figure \ref{fig:SpectEvo}. Both the ions and electrons show a Maxwellian distribution with significant non-thermal populations. In both the ion and electron spectra, the maximum energy grows with time. The electrons equilibrate with a temperature $\approx 0.23$ times the ion temperature.  The power-law of electron distribution is consistent with $f(p)\propto p^{-4.2}$ ($dn/dE\propto E^{-2.2}$). A $p^{-4.2}$ spectrum is predicted for DSA in ultra-relativistic shocks with $\gamma_0\gg1$, while $p^{-4}$ is predicted for non-relativistic shocks \citep[e.g.][]{2001MNRAS.328..393A}. In the ions the downstream spectral index appears softer than the electrons, but the spectral index is not well constrained. A proper measurement of the ion spectrum would require a simulation that was run significantly longer with a larger extent of the non-thermal ion power-law. In our simulation, $\epsilon_p$, the fraction the shock's energy that is put into non-thermal ions is $\sim 0.1$.  The downstream electron temperature is 30\% of the ion temperature, and $\epsilon_e\sim 5\times10^{-4}$. To measure $\epsilon_e$ and $\epsilon_p$ we integrate the non-thermal spectrum starting at a momentum $p_{\rm inj} = 2\gamma_0\beta_0 m_i c$ as measured in the downstream frame.

The upstream spectral evolution with time is shown in Figure \ref{fig:UpSpectEvo}. Both the ions and electrons show two-humped spectra. The lower-energy bump corresponds to the incoming beam of particles while the higher-energy bump is the reflected particles. The ions rapidly converge to having a minor amount of preheating, as lower bump is well fit by a Maxwellian. In contrast, the electrons show strong heating at times $\omega_{pi}t=562$ and 1125, which we attribute to the Weibel filaments which transfer a large fraction of the ions' kinetic energy to the electrons upstream, causing a significant amount of preheating of the electrons, increasing the electron gyroradius \citep[e.g.,][]{2015ApJ...806..165K}. The initial Weibel filaments are effective at injecting electrons not only due to this preheating, but also because the perturbations cause the shock to be highly random in magnetic field obliquity, providing locally quasi-perpendicular field orientation that is conducive to electron injection via shock drift acceleration \citep{2001PASA...18..361B,2015PhRvL.114h5003P}. At later times, after the shock switches to being Bell-mediated, the upstream magnetic fluctuations grow to larger size scales, and the energy transfer between reflected ions and incoming electrons decreases tremendously. This can been seen by the fact that the lower energy electron hump in Figure \ref{fig:UpSpectEvo} becomes increasingly better fit by a simple Maxwellian.

The two species enter the shock front in approximate thermal equilibrium. Since the momenta of the electrons entering the shock are far less than that of the ions, the scale separation between incoming electrons and ions is preserved. There is a minimum energy electrons must attain to escape far upstream, $E_{\rm min}\sim \gamma_0^2 m_i c^2$, visible in the hole in the electron phase diagram, see Figure \ref{fig:BigOne}, and in the time evolution of the upstream spectra, see Figure \ref{fig:UpSpectEvo}. All of the reflected electrons have energies comparable to the reflected ions. As we argue in Section \ref{sec:WideEnough}, the Bell instability creates regions of the shock that are locally superluminal. The size of these regions is comparable to the ion gyro-radius. Electrons need to attain energies comparable to the ion gyro-radius to escape these regions. This hints that the acceleration of electrons in our simulation is likely done first via shock drift acceleration and then DSA as seen in the simulations of \citet{2015PhRvL.114h5003P}.

For both ions and electrons, the highest energy particle of each species has roughly the same kinetic energy. The two species are expected to have the same maximum energy if the rate of energy gain  depends only on the rigidity of the particle and not the charge, like in diffusive shock acceleration. The maximum energy grows linearly with time and does not saturate, $E_{\rm max}\propto t$ (see Figure \ref{fig:Emax_v_t}). The rate at which the maximum energy particle gains energy is consistent with the Bohm limit in standard DSA theory ($E\propto t$, $D\propto E$, where $D$ is the diffusion coefficient). The linear scaling is much faster than $E_{\rm max}\propto \sqrt{t}$ scaling found from small-angle scattering in relativistic Weibel-mediated shocks \citep{2013ApJ...771...54S, 2011A&A...532A..68P}, and makes a large difference the  potential maximum energy particle produced by an astrophysical system. 
\section{Importance of the filamentary mode}
\label{sec:WideEnough}
\begin{figure*}
\centering
\includegraphics[width=0.75\textwidth]{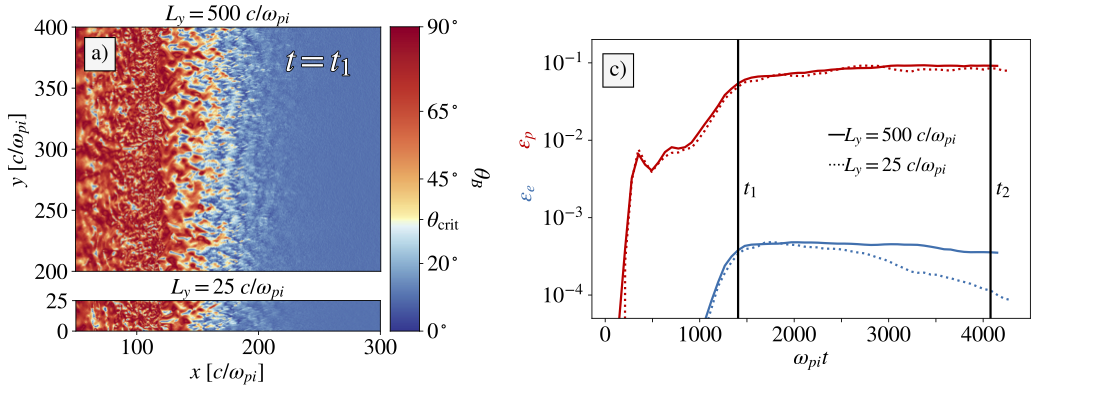}
\includegraphics[width=0.75\textwidth]{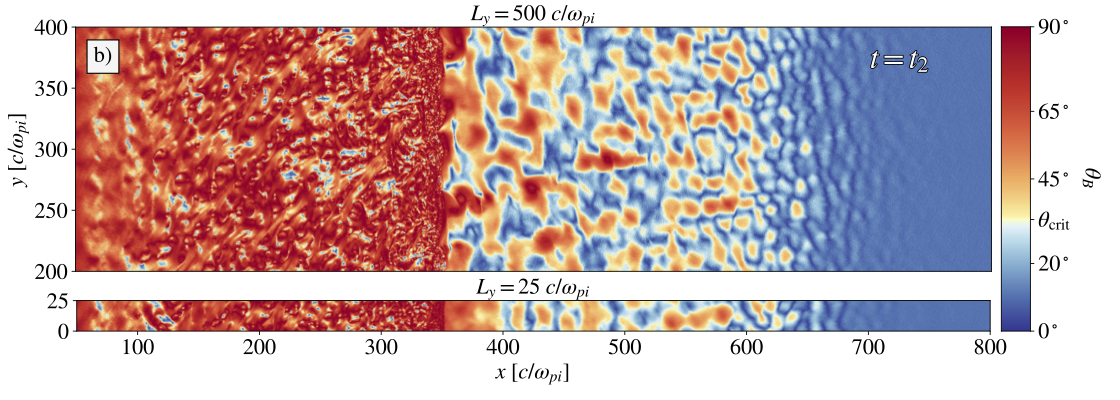}
\caption{Magnetic field structure in the fiducial simulation (width $L_y = 500\ c/\omega_{pi}$) is compared to a simulation with a smaller transverse size ($L_y = 25\ c/\omega_{pi}$). We show the angle of the magnetic field with the shock normal, $\theta_B\equiv\tan^{-1}{|B_\perp/B_\parallel|}$ for two simulations at two times: $t_1 = 1400/\omega_{pi}$, when the shock begins to become Bell mediated, and $t_2 = 4078/\omega_{pi}$. The angle at which the shock transitions between subluminal and superluminal, $\theta_{\rm crit}$, is marked on the colorbar. The larger simulation is shown as the upper panel in a) and b), and is plotted with the solid lines in c). The smaller $L_y = 25\ c/\omega_{pi}$ simulation is shown in the lower panels of a) and b) in the 2D plots and dotted lines in c). At earlier times, the two simulations agree. However, at late times, the shock front has patches of superluminal and subluminal magnetic field orientations in the fiducial run, while in the run with a smaller $L_y$, the shock is superluminal across the entire transverse direction. The effect of the additional magnetic turbulence on particle acceleration is shown in panel c). When the transverse waves fill the box in the $L_y = 25\ c/\omega_{pi}$ run, $\epsilon_e$ decreases.
}
\label{fig:Width}
\end{figure*}
We find that to properly recover the correct shock structure, and measure $\epsilon_e$, the simulation domain must be wide enough so that the transverse waves cannot fill the entire box. This criterion is satisfied when the size of the box is much larger than the gyroradius of the ions carrying the majority of the current upstream. When using a transverse size that is too small, the Bell modes can grow until they fill the transverse size. When this happens, $\epsilon_e$ is suppressed. The characteristic angle of the amplified magnetic field, $\theta_B \equiv \tan^{-1}{|B_\perp/B_\parallel|}$, is $\gtrsim 45^\circ$. For $\beta_s =0.83$, the angle at which the shock becomes superluminal is $\theta_{\rm crit} = \cos^{-1}{\beta_s} \approx 34^\circ$, where $\theta$ is measured in the upstream frame \citep{1990ApJ...353...66B,2011ApJ...726...75S}. The existence of locally quasi-parallel regions of the shock front allow electrons to escape back upstream.

The size needed to properly resolve Bell cosmic ray current driven instability depends on the size of the evacuated cavities upstream. The cavities grow until they are advected towards the shock. The size of the cavities when they impact the shock front depends on the ratio of the advection time to the growth rate of the cavities \citep[\(\Gamma\propto |B_\perp|\xi^{1/2}_{\rm acc}\), where $\xi_{\rm acc}$ is the pressure of the accelerated ions,][]{2012MNRAS.419.2433R, 2013ApJ...765L..20C}.  The maximum size is ultimately limited by the gyroradii of the ions carrying the current. The gyro-radius of a typical shock reflected ion ($E\sim\gamma_0^2m_i c^2$) is $r_{gyr}\sim 30\ c/\omega_{pi}$. \citet{2013ApJ...765L..20C} found that in hybrid simulations of non-relativistic shocks, the size of the filament can grow to be comparable to the gyroradius of the highest energy ion.

To show the importance of the transverse size of simulation for the measurement of $\epsilon_e$, we compare our fiducial run that is much wider than the typical ion gyro-radius to a simulation with a width slightly smaller than the ion gyroradius in Figure \ref{fig:Width}. At early times, the smaller, $L_y = 25\ c/\omega_{pi}$, run agrees with the fiducial, $L_y = 500\ c/\omega_{pi}$, run in terms of magnetic structure (Panel a) and particle acceleration (Panel c). However, as we can see in Panel b, failing to resolve the ion gyroradii significantly changes the shock structure, where the fiducial run shows filaments with width $\sim r_{gyr}$, but the smaller run does not have filaments at the shock front. The filaments in the fiducial run cause the shock front to have regions of subluminal and superluminal magnetic field orientations at the same time. Without filaments, as in the smaller run, alternating strips of subluminal and superluminal fields fill the transverse direction upstream. This difference in structure changes $\epsilon_e$ (see Panel c of Figure \ref{fig:Width}). The box must be sufficiently wide to capture the filaments to properly measure the energy in non-thermal electrons in mildly relativistic shocks. 

Electron injection into shock acceleration decreases if the transverse size of the waves can grow to the entire box. We confirmed that the decrease in $\epsilon_e$ is not just a matter of statistics by running small runs with larger ppc (up to 64).  We also compared our fiducial, $L_y  = 500\ c/\omega_{pi}$, run to slightly smaller run that still captured the gyro-radius, $L_y  = 200\ c/\omega_{pi}$. We saw no difference in $\epsilon_e$ in those runs, suggesting it is sufficient to capture a few patches of a locally quasi-parallel shocks to properly measure $\epsilon_e$.
\section{Quasi-perpendicular Mildly Relativistic Shocks}
\label{sec:SuperLum}
\begin{figure}
\centering
\includegraphics[width=\linewidth]{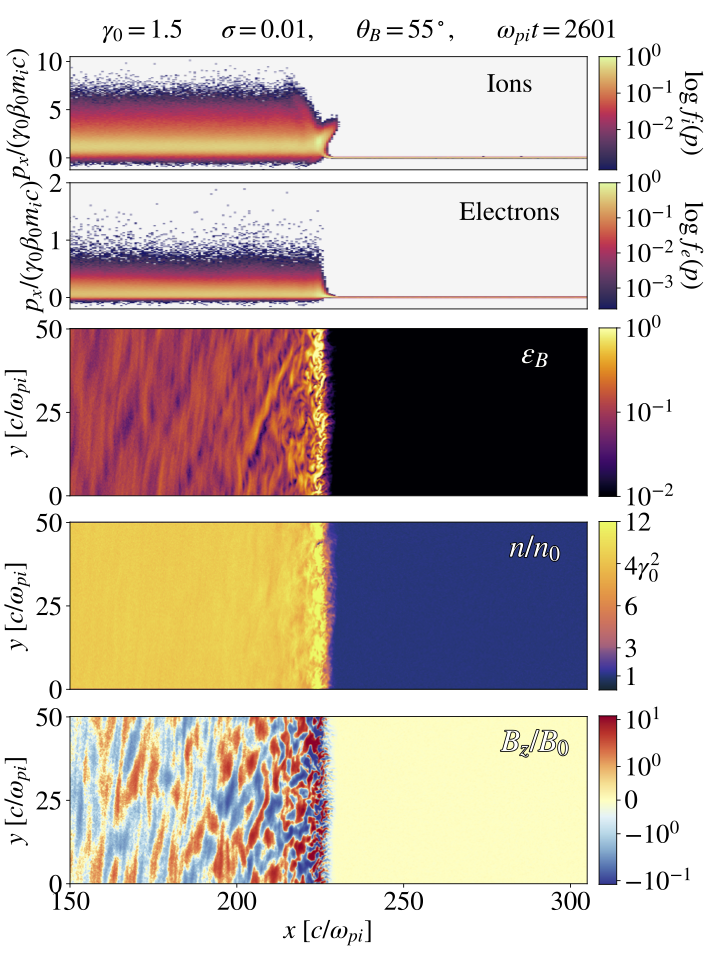}
\caption{Structure of a superluminal mildly relativistic shock with $\theta_B = 55^\circ$. Compare this figure to the subluminal case with $\theta_B\approx 10^\circ$ in Figure \ref{fig:BigOne}. There are no particles escaping upstream and no upstream turbulence, as expected for superluminal magnetized shocks.}
\label{fig:ObliqueFields}
\end{figure}

\begin{figure}
\centering
\includegraphics[width=\linewidth]{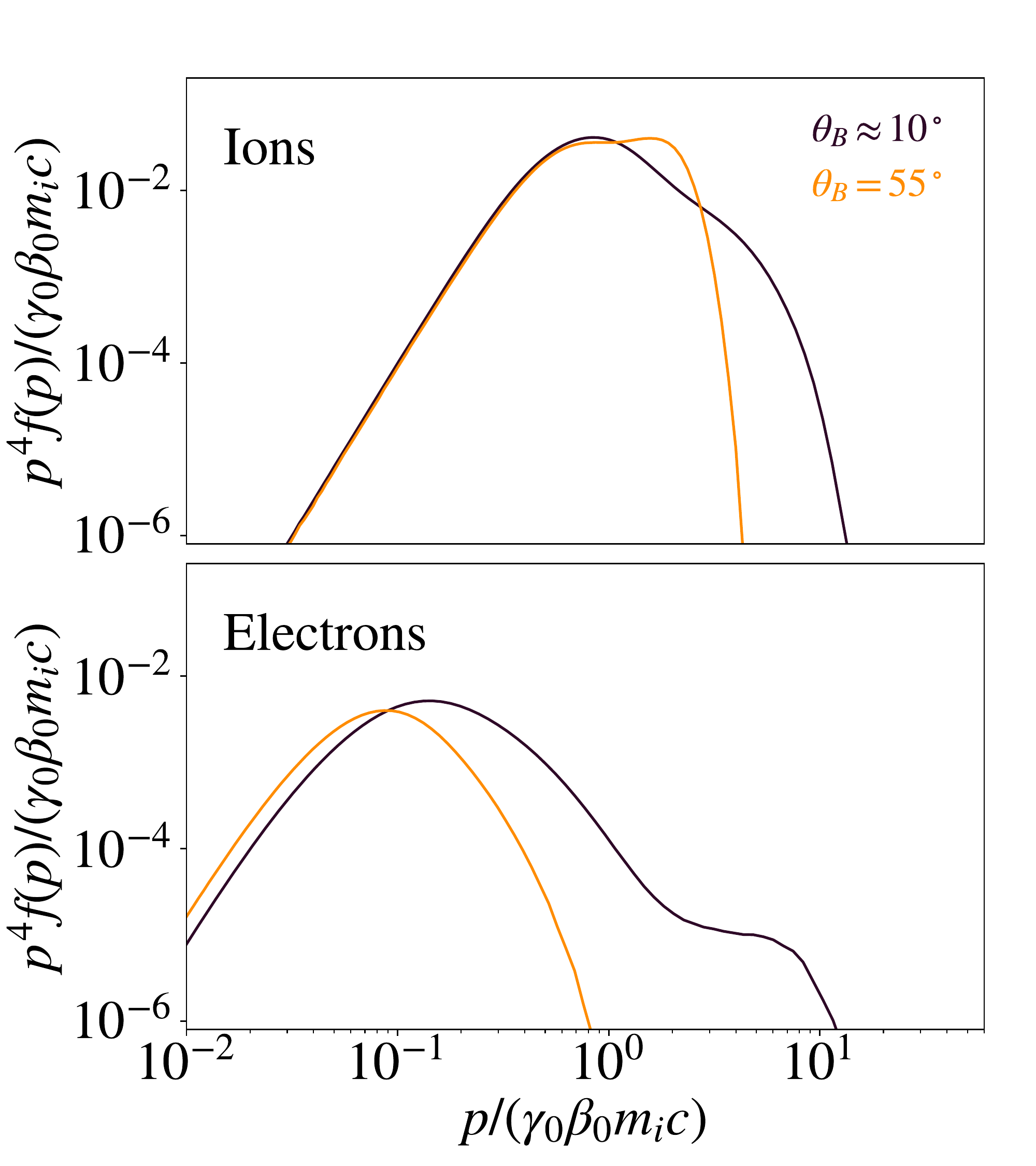}
\caption{A comparison of the downstream spectra of a superluminal shock with $\theta_B =55^\circ$ and the subluminal shock with $\theta_B\approx 10^\circ$ at time $\omega_{pi}t\approx 2600$ measured 10-80$c/\omega_{pi}$ downstream from the shock. In superluminal shocks, neither electrons nor ions show non-thermal power laws, but the ion spectrum does show a bump from shock-drift acceleration.}
\label{fig:SubVSup}
\end{figure}

Here we change the magnetic field orientation to examine a quasi-perpendicular mildly-relativistic shock with $\theta_B=55^\circ$. For $\gamma_0=1.5$, the angle at which the shock becomes superluminal is $\theta_{\rm crit} \approx 34^\circ$. Therefore with $\theta_B=55^\circ$, particles are unable to escape upstream and drive turbulence far from the shock. The inability of particles to escape far upstream prevents them from being injected into DSA and forming a non-thermal power-law. 

In Figure \ref{fig:ObliqueFields}, we show an example of a superluminal, quasi-perpendicular shock. Unlike the quasi-parallel shock, there is no turbulence upstream and little turbulence downstream. The lack of upstream magnetic turbulence results in weaker magnetic fields downstream. 

In the quasi-perpendicular shock, a prominent suprathermal bump from shock drift acceleration (SDA) is visible in the downstream ion spectra shown in Figure \ref{fig:SubVSup}. However, the maximum energy in both species does not grow with time or develop into a power-law. Furthermore, no particles are reflected far upstream from the shock front. The lack of reflected upstream particles means that there are no particles to seed magnetic turbulence upstream. Therefore, there is no preheating upstream; $T_p=T_e=T_0$. Downstream, $T_p \approx 6 T_e$. The ion temperature is nearly the same in the subluminal and superluminal shocks, but $T_e$ of the superluminal shock is about half of $T_e$ in the quasi-parallel case.  The additional turbulence provided by the shock-reflected particles in quasi-parallel shocks drives the downstream species closer to equilibrium.

As we show in appendix \ref{appendix:Bplane}, when $\theta_B\sim10^\circ$, the shock structure and particle acceleration do not depend on the orientation of the magnetic field with respect to the simulation plane. However in quasi-perpendicular shocks, the turbulence right at the shock front and downstream depends strongly on the orientation of the field with respect to the simulation plane. In the out-of-plane configuration, ions gyrate around $B_z$ in the $x-y$ plane, reducing the effective adiabatic index, causing downstream magnetic fluctuations, and decreasing the importance of SDA in the ions. But for both in-plane and out-of-plane superluminal magnetic field configurations, particles do not return upstream and a non-thermal power-law does not develop. 

\section{Conclusions}
\label{sec:Conclusions}
This paper is the first in a series that examines how particle acceleration in shocks changes when the shock transitions from non-relativistic to relativistic. We used large computational domains and a fully kinetic method. We captured the filamentary behavior of the Bell instability. In doing so, we have followed the shock from formation to the shock's later stages, resolving all kinetic ion and electron scales. In this paper, we have shown that properly capturing the non-resonant instability is crucial for measuring the fraction of the shock's energy that goes into non-thermal electrons, $\epsilon_e$.

There is observational and theoretical evidence that both relativistic and non-relativistic quasi-parallel shocks are efficient ion accelerators with a $\epsilon_p\sim 0.1$. We find that the fraction of the shock's energy in non-thermal electrons, $\epsilon_e$, to be $\sim5\times10^{-4}$ for a mildly relativistic, magnetized shock with $\theta_B=10^\circ$, $\gamma_0 =1.5$, $\beta_0\approx0.75$, and $M_\mathcal{A}\approx15$. Our measurement of $\epsilon_e$ can be directly compared to other PIC simulations in the literature of shocks with $\sigma\sim0.01$ or Alfv{\'e}nic Mach numbers $\sim 10$. Our reported $\epsilon_e$ is consistent with than the 1D measurement of $\epsilon_e\sim {\rm a \ few} \times10^{-4}$ for non-relativistic shocks with $\beta_0 = 0.1$ by \citet{2015PhRvL.114h5003P}, and less than the $\epsilon_e\sim0.1$ measured in relativistic shocks by \citet{2011ApJ...726...75S}. Our simulations are also consistent with the trend of increasing $\epsilon_e$ as the shock becomes more relativistic, but a set of directly comparable simulations spanning the non-relativistic to relativistic regime is the goal of a future work.

Our simulations had a moderately large magnetization $\sigma_0 = 0.007,\ M_\mathcal{A}\approx M_s \approx 15$: a setup consistent with possible internal shocks in the jets of microquasars or quasars. We show that magnetized, subluminal, mildly-relativistic shocks are capable of producing non-thermal ions and electrons in hard power-laws whose maximum energy grows linearly with time. 

The evolution of a quasi-parallel weakly magnetized shock can be summarized as follows: the shock forms via the Weibel instability which starts the Fermi process and reflects particles into the upstream. The Weibel instability efficiently preheats incoming electrons to an energy comparable to the ions. Therefore, both electrons and ions are efficiently injected in a Weibel-mediated shock. However, the Weibel instability only efficiently scatters particles back towards the shock until a certain energy. Above this energy the gyroradii of the particles are too large to be effectively scattered, and the high-energy particles can freely stream far away from the shock due to the quasi-parallel field geometry. The escaping high-energy ions drive a current in the upstream plasma, which grows transverse waves via the Bell cosmic-ray filamentary instability.  After about $\sim 25$ growth times, the Bell instability becomes the dominant instability that mediates the shock. Once mediated by the Bell filamentary instability, the shock and non-thermal particles change in the following ways:
\begin{enumerate}
\item For $\mathcal{M_A}=15$, there is large magnetic field amplification both upstream and downstream. The y-averaged field amplification upstream is as $\sim 2B_0$, and close to the shock the most amplified field can be $\gtrsim 5B_0$. The largest upstream field amplification occurs at the walls of worm-like cavities evacuated by energetic ions. A $\mathbf{J}\times \delta \mathbf{B}$ force acts on a current carried by the background plasma, which is balancing the current of high energy ions that escape far upstream. The force evacuates these worm-like cavities. The transverse size and  extent of these structures grows with time. Downstream, $\epsilon_B$ is both larger and extends further away from the shock than when the shock is Weibel-mediated (see Figures \ref{fig:epsBvt} \& \ref{fig:AppendixBepsB}).

\item The shock structure is highly turbulent and magnetic field amplification downstream happens in two structures: long magnetized filaments with a characteristic $\epsilon_B\gtrsim0.1$ and compact underdense holes with a large out-of-plane magnetic field and an $\epsilon_B\sim1$, confirming the turbulent structure found by \citet{2013ApJ...765L..20C} using hybrid simulations. When an upstream cavity is advected downstream, it forms two underdense holes of opposite $B_z$ sign. In 3D they would likely form a connected twisted flux tube. The holes last for a long time and even merge in 2D. They may be an important place for magnetic dissipation, but the dynamics needs to be checked in 3D.

\item The filamentary nature of the Bell instability plays a vital role in electron injection in mildly relativistic shocks. At early times, the Weibel instability efficiently reflects both ions and electrons at the shock front, as in relativistic shocks \citep[e.g.,][]{2013ApJ...771...54S}. After the shock becomes Bell-mediated, $\omega_{pi}t\sim 1500$ in our fiducial run, electron reflection and injection into DSA becomes less effective. If the simulation box is not wide enough to capture the gyro-radius of the non-thermal ions, $\epsilon_e$ will be suppressed significantly in shocks with $\gamma\beta\sim 1$. 

\item The maximum energy of both the ions and electrons in the quasi-parallel simulation increases linearly after the shock becomes Bell-mediated. The maximum ion and electron energy is the same. The linear increase in energy is consistent with the Bohm scaling, $E_{\rm max}\propto t$, and is much faster than the scaling of $E_{\rm max}\propto t^{1/2}$ in Weibel-mediated, high-Mach number, ultra-relativistic shocks \citep{2013ApJ...771...54S}. Because $E_{\rm max}$ grows much faster in mildly relativistic shocks, sources with mildly relativistic shocks like FRI type AGN may be more likely progenitors of ultra-high energy cosmic rays than the ultra-relativistic unmagnetized shocks in GRBs.

\item Electrons are not preheated to a large fraction of the reflected ion energy in trans-relativistic shocks. In relavistic and mildly relativistic Weibel mediated shocks, the reflected ions are capable of transferring a sizeable portion of their energy to the electrons upstream, so that the rigidity of the two species is similar at the shock front. We find that once the shock becomes Bell-mediated, ions and electrons are preheated by similar amounts; at one ion gyro-radius in front of the shock, the two temperatures are $T_e\sim3T_0$ and $T_p\sim 4T_0$, where $T_0$ is the initialized temperature of the plasma. Therefore, the two species enter the shock region in approximate thermal equilibrium, with thermal momenta far less than the bulk momentum of an ion. Thus, the large difference between ion and electron rigidity at the shock front is preserved. Downstream and right at the shock front, the ions do transfer their energy to the electrons, and the downstream electrons equilibrate with a temperature $\sim 30\%$ of the ions' downstream temperature.

\item To be reflected upstream, the electrons must have an energy comparable to a shock-reflected ion, meaning electrons must gain a lot of energy close to the shock front, possibly through a shock drift acceleration (SDA) process as seen in 1D simulations of Bell's mediated electron-ion shocks \citep{2015PhRvL.114h5003P}, but we did not study the orbits of the non-thermal electrons. We leave that to a future work.

\item Magnetized, quasi-parallel shocks are efficient ion accelerators. Approximately $10\%$ of the shock's energy goes into non-thermal ions. The power-law index of ions appears to be softer than $f(p)\propto p^{-4}$ ($dn/dE\propto E^{-2}$), but the spectral index is not well constrained by our simulation. Constraining the ion spectrum would require much longer simulations.

\item Quasi-parallel shocks are capable of injecting electrons into a hard power-law with a spectral index consistent with $f(p)\propto p^{-4.2}$ ($dn/dE\propto E^{-2.2}$). The power-law starts at $\sim 2\gamma_0\beta_0 m_i c$.  For $\mathcal{M_A}=15$, $\theta_B = 10^\circ$, $\gamma_0 = 1.5, \beta_0\approx 0.75$, the amount of the shock's energy that is put into the non-thermal electron power-law is small, $\epsilon_e\sim 5\times 10^{-4}$. This $\epsilon_e$ corresponds to a number ratio of non-thermal electrons to total electrons of $5\times 10^{-5}$.  However, the value of $\epsilon_e$ may depend on the magnetic inclination upstream, $\sigma_0$, and the speed of the shock. We will address these trends in a future work.
\end{enumerate}
Finally, we examined particle acceleration in superluminal mildly relativistic shocks, with $\theta_B=55^\circ$. Since particles were unable to escape upstream, neither species was capable of entering DSA. Instead, a fraction of ions gained energy via shock drift acceleration while crossing the shock front, resulting in a second suprathermal bump in the downstream ion spectrum. In mildly relativistic superluminal shocks, the maximum energy of both species did not increase with time.
Throughout this paper we have used a reduced mass ratio of $m_i/m_e=64$. We find that $m_i/m_e$ must be fairly large, or else Bell instability will be suppressed. We also find that 64 is sufficiently large to capture the shock structure and $\epsilon_e$, as shown in Appendix \ref{sec:MassRatio}.

Our results on acceleration in mildly relativistic shocks have the following implications for simulations of electron-ion shocks: a small reduced mass ratio, numerical heating, and too short of simulation duration all work in concert to suppress the Bell instability. If Bell does become the main instability, the simulation must have a large transverse size to capture the upstream filaments, and these filaments play a big role in injecting electrons in mildly relativistic shocks. These conclusions are very general and likely apply equally well to non-relativistic and relativistic particle-in-cell simulations of quasi-parallel shocks. 

For non-relativistic shocks traveling at $v\ll c$, the filamentary mode of the Bell instability may not be as important for acceleration of electrons because even the amplified magnetic field is subluminal. This may explain the close agreement between $\epsilon_e$ measured in this work and the 1D simulations of $v=0.05c$ shocks in \citet{2015PhRvL.114h5003P}. A set of directly comparable, 2D simulation of shocks ranging from non-relativistic to relativistic is left to future work.

Since our shocks were mildly relativistic, one needs to apply caution when extrapolating our conclusions regarding Bell mediated shocks to  relativistic or non-relativistic collisionless electron-ion quasi-parallel shocks. The growth rates of the microphysical instabilities may depend on physical parameters like magnetization and shock velocities. If $\sigma_0$ is large, the shock is likely mediated by the resonant ion instability, and when  $\sigma_0 = 0$ or is very small, there is not enough magnetic field on large scales to initiate Bell instability. If the field orientation is quasi-perpendicular with little upstream turbulence, there may not be enough ion injection to cause the Bell instability. We will address how electron acceleration efficiency and magnetic amplification depends on $v/c$ and $\sigma_0$, constraining the models that assume equipartition between $\epsilon_B$ and $\epsilon_e$ in future work.

In the future, it will be important to determine if 2D PIC simulations can properly capture the physics of a Bell-mediated shock. The reduced dimensionality of the simulations may allow for structures that do not form in 3D. A 3D simulation will be expensive because to be fully 3D the $z$ direction should be much larger that the typical ion gyroradius and run for a long time. However, we can use the 3D results of prior hybrid simulations to inform how PIC simulations might change if run in 3D. \citet{2014ApJ...783...91C} found that in 3D the reflected cosmic rays still evacuated cavities upstream, and the magnetic field amplification and ion injection and diffusion was largely the same. Therefore, our general picture of a transition from a Weibel- to a Bell-mediated quasi-parallel shock with an $\epsilon_p\sim10\%$, $E_{\rm max}\propto t$ should hold. The characteristic values for the magnetic field amplification upstream and the less-amplified magnetic filaments downstream would likely remain in a 3D PIC simulation as well. However, our simulations have features not capable of being captured in hybrid simulations---the electron preheating of a factor $\sim 3 T_0$, and $\epsilon_e\sim5\times10^{-4}$. A fully 3D PIC simulation would need to be run to answer how our results may change.

In conclusion, we have used fully kinetic particle-in-cell simulations to show that quasi-parallel, magnetized, mildly-relativistic electron-ion shocks traveling at velocity $v\sim 0.7c$  are capable of self-consistently generating the necessary magnetic turbulence upstream to accelerate particles to high energies quickly. In addition, these shocks put a large fraction of their energy into hadrons. These characteristics suggest that astrophysical objects that host such shocks may be an important source of high-energy hadronic radiation, cosmogenic neutrinos, and possibly a source of the ultra-high energy cosmic rays. Therefore, objects like low-luminosity AGN like BL Lacs and FRI, as well as micro-quasars, which may have mildly-relativistic shocks inside of their jets, are important potential multi-messenger sources for the upcoming high-energy $\gamma$-ray telescopes such as the Cherenkov Telescope Array and also existing and forth-coming high-energy neutrino detectors like IceCube and KM3NeT.

\section*{Acknowledgements}
We thank Jaehong Park and Lorenzo Sironi for valuable discussions and their assistance with {\it TRISTAN-MP}.
PC acknowledges financial support from the WARP program of the Netherlands Organisation for Scientific Research (NWO). SM is supported by an NWO VICI grant (no. 639.043.513). 
PC thanks Rahul Kumar and Illya Plotnikov for valuable conversations. The simulations in this paper were performed using computational resources at the TIGRESS high performance computer center at Princeton University, the University of Chicago Research Computing Center, and the Texas Advanced Computing Center (TACC) at The University of Texas at Austin (TG-AST100035). This work was supported by NSF grant AST-1517638. AS is supported by the Simons Foundation (grant 267233).

\bibliographystyle{mnras}
\bibliography{particle_acceleration.bib}

\appendix

\section{Scaling with mass ratio}
\label{sec:MassRatio}
\begin{figure}
\centering
\includegraphics[width=0.85\linewidth]{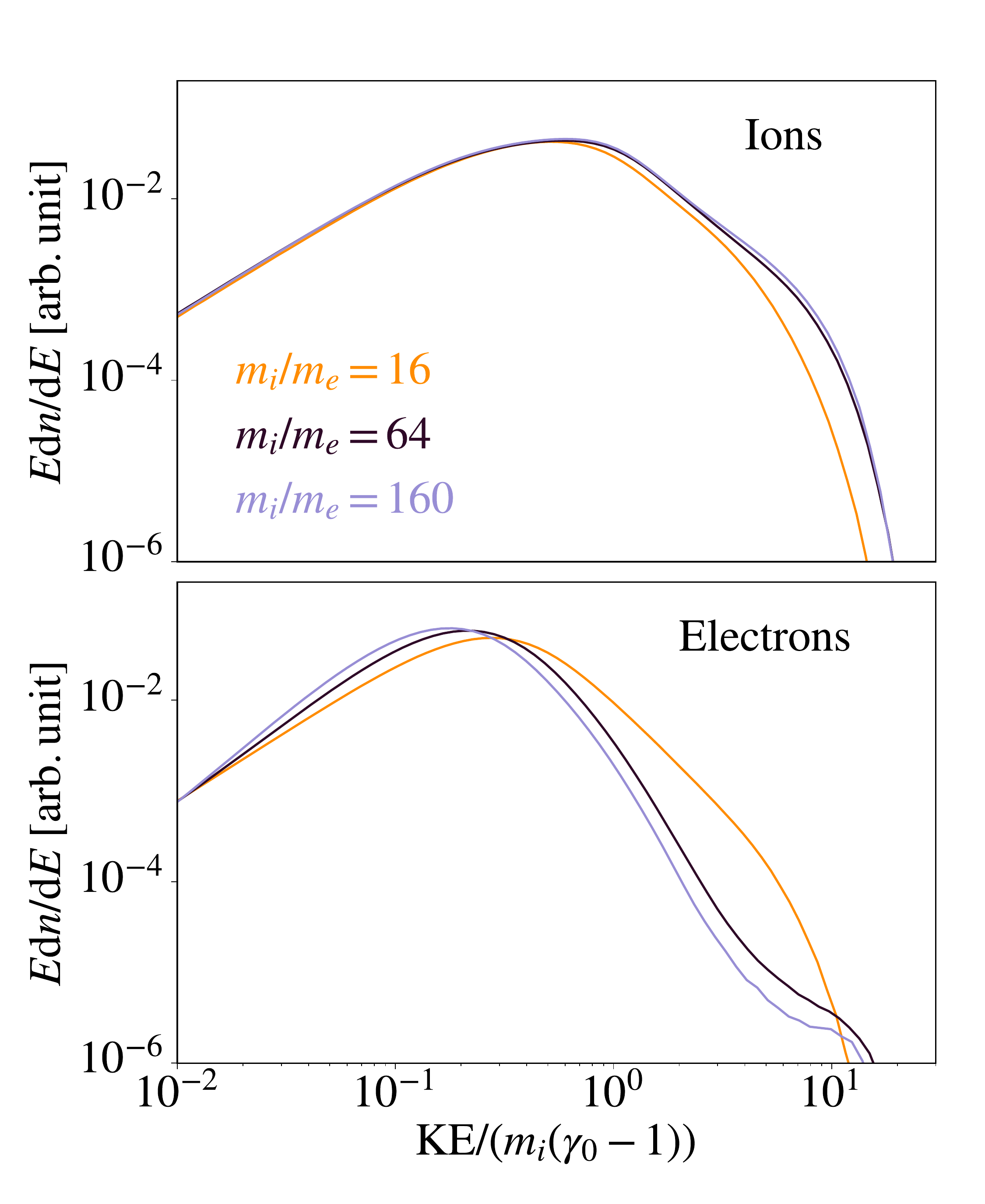}
\caption{The two panels show the downstream particle spectra in the downstream rest frame extracted 10--80 $c/\omega_{pi}$ downstream of the shock at $\omega_{pi}t \approx 2000$, shortly after the time the the shock becomes Bell-mediated in the fiducial, $m_i/m_e = 64$, simulation. As the mass ratio increases to more realistic values, the energy transfer to electrons from ions decreases, with electrons having a lower temperature downstream. For $m_i/m_e =16$, the cosmic ray filamentary instability is significantly suppressed and $\epsilon_e$ is far too large at $\omega_{pi}t\sim2000$.}
\label{fig:AppendixBDownSpect}
\end{figure}
\begin{figure}
\centering
\includegraphics[width=0.85\linewidth]{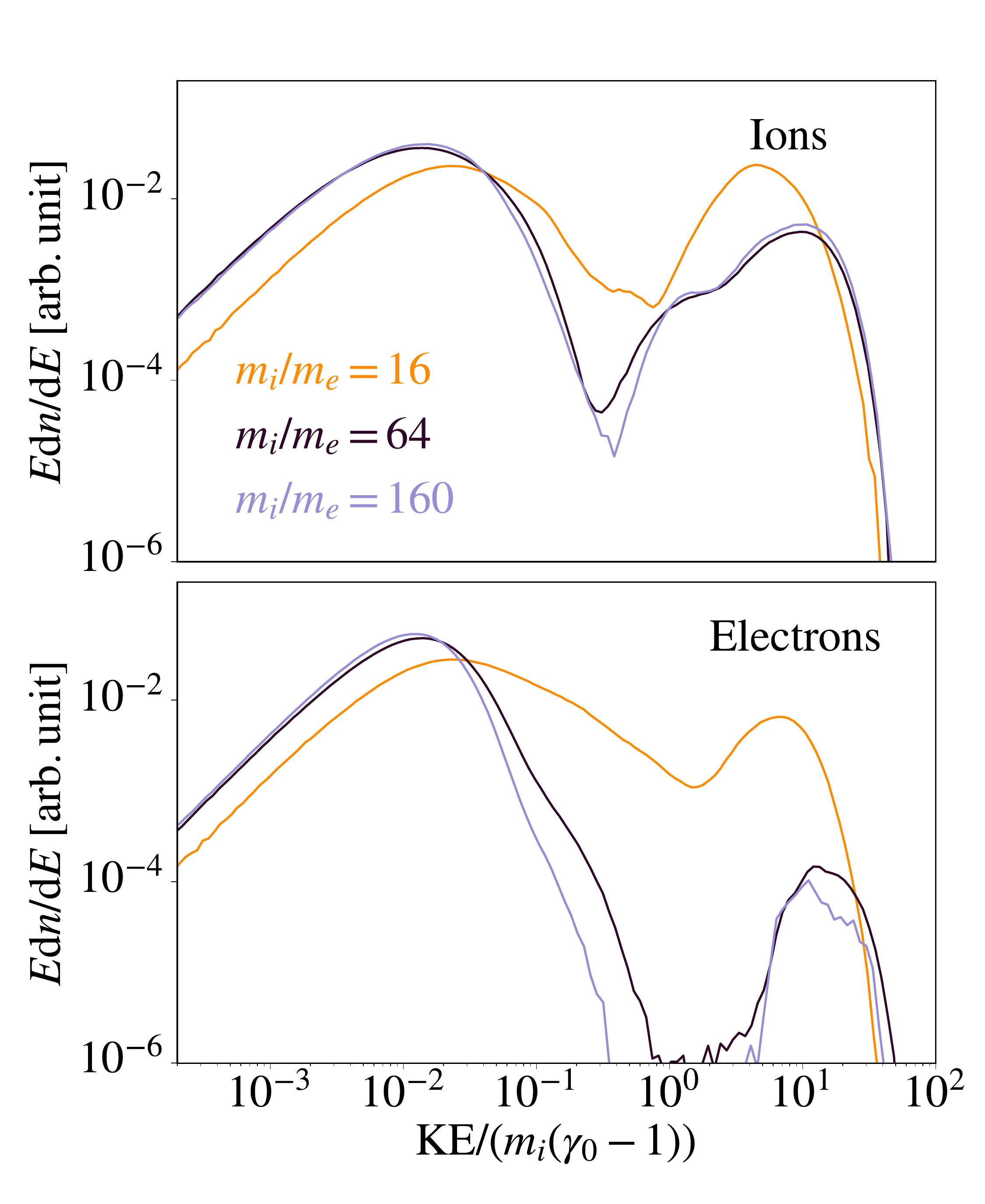}
\caption{Three simulations that illustrate the importance of a large mass ratio. The two panels show the particle spectra in the upstream rest frame extracted 10--20 $c/\omega_{pi}$ upstream of the shock at $\omega_{pi}t \approx 2000$. With a small of mass ratio, electrons are heated significantly more upstream before entering the shock.}
\label{fig:AppendixBUpSpect}
\end{figure}

\begin{figure}
\centering
\includegraphics[width=1
\linewidth]{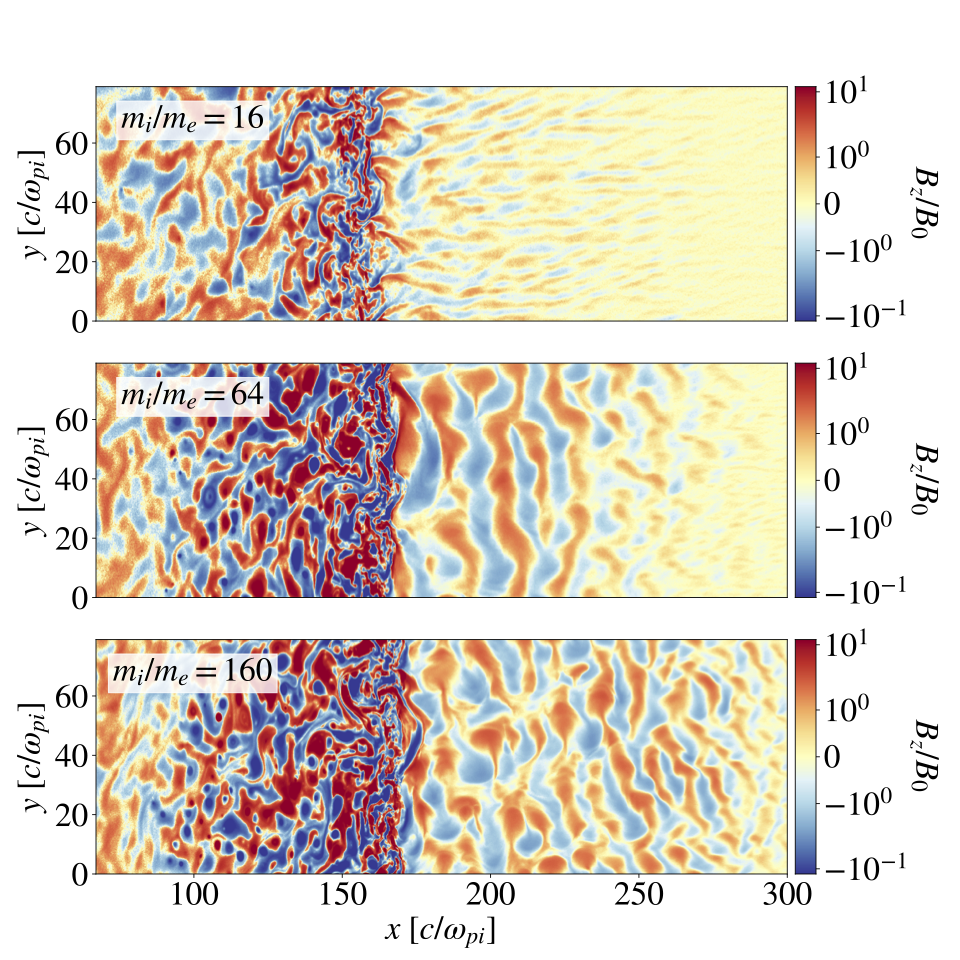}
\caption{The effect of varying the mass ratio on the filamentary structures in the upstream. The three panels show $B_z$  near the shock front at time $\omega_{pi}t \approx 2000$. The middle panel shows a representative slice of the fiducial simulation. Increasing $m_i/m_e$ increases the importance of the cosmic ray filamentation instability. }
\label{fig:AppendixBDens}
\end{figure}

\begin{figure}
\centering
\includegraphics[width=1\linewidth]{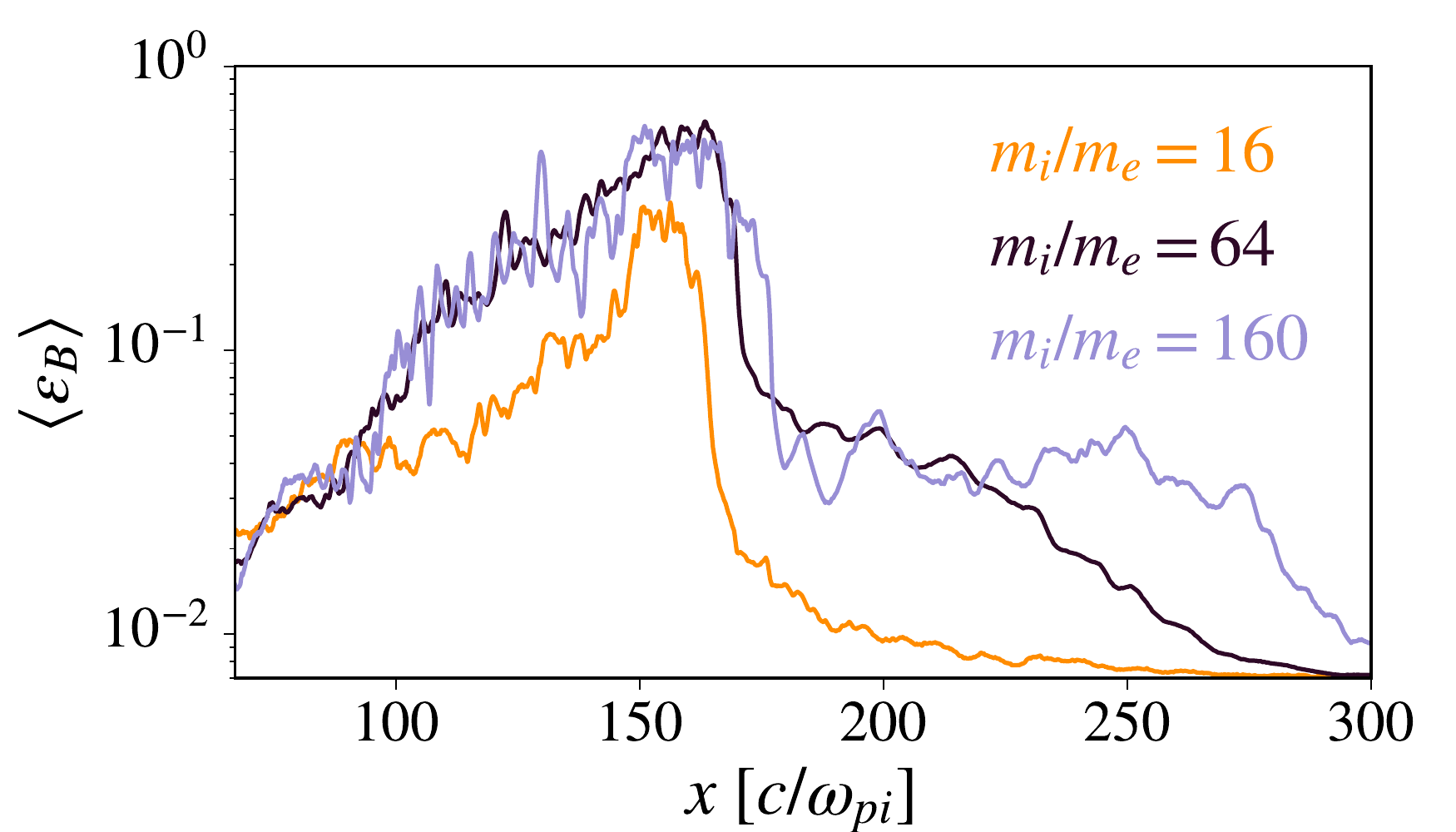}
\caption{The effect of varying the mass ratio on the $y$-averaged $\epsilon_B$ at time $\omega_{pi}t \approx 2000$. When more realistic mass ratios are used, the cosmic ray filamentation instability becomes more important, increasing the magnetic field amplification both upstream and downstream of the shock front. Note the good quantitative agreement between $m_i/m_e =64$ and 160 in terms of the magnitude of the magnetic amplification upstream and downstream.}
\label{fig:AppendixBepsB}
\end{figure}
In the main text of paper we have employed a reduced mass ratio $m_i/m_e=64\ll 1836$. Here we examine the consequences of using a smaller mass ratio on the shock structure and particle acceleration. We find that using too small of a mass ratio can suppress the cosmic ray filamentation instability with the following implications---there is too little magnetic field amplification upstream and downstream, the maximum particle energy grows too slowly, the energy transfer between electrons and ions is too efficient, and artificially small $m_i/m_e$ boosts $\epsilon_e$. While our fiducial mass ratio of $m_i/m_e = 64$ agrees quantitatively with a higher mass ratio $m_i/m_e = 160$ simulation in terms of the magnetic field amplification and maximum particle energy, it only qualitatively agrees in the other respects. Using $m_i/m_e =64$ slightly under-predicts the importance of the cosmic ray filamentation instability, with smaller, weaker filaments in the upstream, more electron heating, and an $\epsilon_e$ that is slightly bigger in the $m_i/m_e = 64$ simulation than $m_i/m_e =160$. In the larger mass ratio simulation whistler waves may be more important, as they will cross the $\mathcal{M_A}\lesssim \sqrt{m_i/m_e}$ barrier at smaller values of magnetic field amplification.

To see the effect of our choice of reduced mass ratio, 64, we compare our fiducial run to a simulation with a smaller mass ratio of 16 \citep[the mass ratio used in][]{2011ApJ...726...75S}, and a larger mass ratio of 160 in Figures \ref{fig:AppendixBDownSpect}, \ref{fig:AppendixBUpSpect}, \ref{fig:AppendixBDens} and \ref{fig:AppendixBepsB}. The higher mass ratio run had the same setup as the fiducial run except it used a smaller computational box with a width of $\approx 79\ c/\omega_{pi}$. The $m_i/m_e =16$ simulation had the same setup except it used a higher resolution of $c/\omega_{pe} = 8$ cells and a box width of $\approx 109 \ c/\omega_{pi}$. The higher resolution was required to ensure that the smaller Debye length was resolved ($\lambda_D\propto\sqrt{m_i/m_e}$).

Figures \ref{fig:AppendixBDownSpect} and \ref{fig:AppendixBUpSpect} show the effect of the reduced mass ratio on the downstream and upstream spectral properties respectively. It is clear that using a mass ratio of 16 is insufficient to correctly capture the spectral properties of a Bell-mediated, electron-ion shock at time $\omega_{pi}t\sim 2000$ in a $\gamma_0=1.5$ shock. However, \citet{2011ApJ...726...75S} do see circularly polarized waves in their simulations of relativistic magnetized shocks with $m_i/m_e=16$, $\sigma=0.1$, and $\gamma_0=15$. The difference may be an issue of relativistic shocks evolving faster, or the higher magnetization ($\sigma=0.1$) used in those simulations.

The $m_i/m_e =16$ run predicts an $\epsilon_e$ that is $> 10$ times larger than the runs with larger mass ratios, whereas there is good agreement between the $m_i/m_e =64$ and 160 runs. There is no sign of an increase in energy transfer in the high mass run due to whistler waves. As the mass ratio increases, the upstream energy exchange between electrons and ions is less efficient. 

The downstream ion spectrum in the $m_i/m_e = 16$ run does not extend to as high of energy as the larger mass ratio runs, so $\epsilon_p$ is not well-constrained, but it appears similar to the larger mass ratio runs. Note that the non-thermal tail extends to larger energies in both the electrons and ions for the larger mass ratio runs as compared to the $m_i/m_e=16$. The high energy particles are scattered back to the shock more efficiently because the larger mass ratio runs show more upstream magnetic field amplification and structures at larger scales (see Figures \ref{fig:AppendixBDens} \& \ref{fig:AppendixBepsB}). 

In terms of the overall magnetic field amplification, the $m_i/m_e=160$ and the fiducial run agree very well both upstream and downstream, the only difference is that the extent of the amplification is larger upstream for $m_i/m_e=160$. The difference in extent is because the larger mass ratio run transitions from Weibel-mediated to Bell earlier than the fiducial run (in terms of $\omega_{pi}^{-1}$). The smaller mass ratio of $m_i/m_e = 16$ has not yet transitioned to being Bell-mediated.
\section{Dependence on Magnetic Field Orientation with respect to Simulation Plane}
\label{appendix:Bplane}
\begin{figure}
\centering
\includegraphics[width=0.45\textwidth]{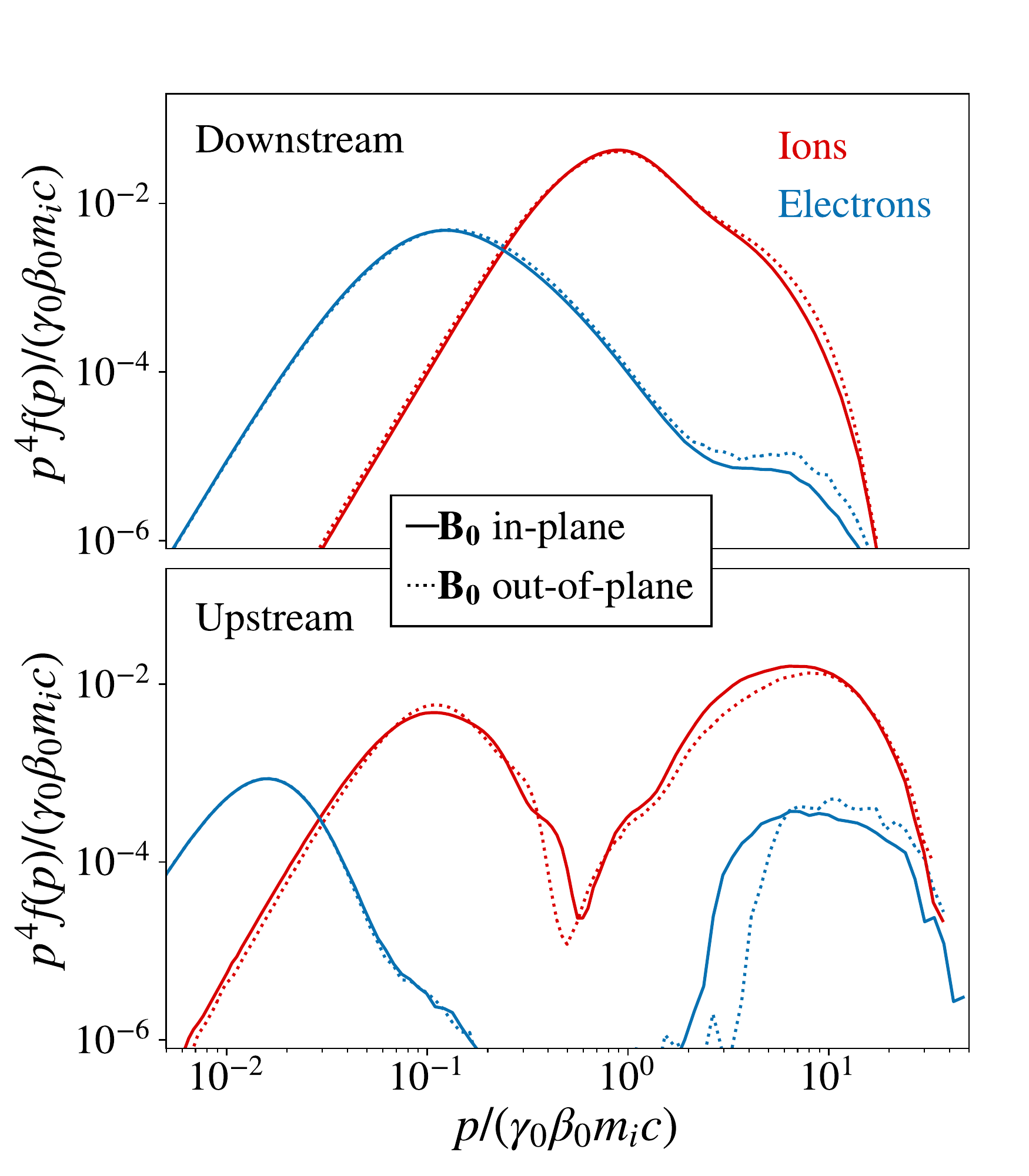}
\caption{Comparison of upstream and downstream spectra with different initial magnetic field orientations with respect to the simulation plane at time $\omega_{pi}t =4148$. Solid lines correspond to the fiducial run with an in-plane $B_0$ field (i.e. $B_{0,\perp}$ along $y$). Dotted lines correspond to a smaller run with out-of-plane orientation ($B_{0,\perp}$ along $z$) and a width of 100 $c/\omega_{pi}$. The upper panel is the particle spectra in the simulation rest frame extracted 20--30 $c/\omega_{pi}$ upstream of the shock, while the lower panel is the spectra in the downstream rest frame extracted 10--80 $c/\omega_{pi}$ downstream of the shock. While the out-of-plane inclination initially reflects more electrons, when turbulence $B$ field upstream grows larger, the difference between the two inclinations electron acceleration efficiency is small.
}
\label{fig:InVOut}
\end{figure}

\begin{figure}
\centering
\includegraphics[width=0.45\textwidth]{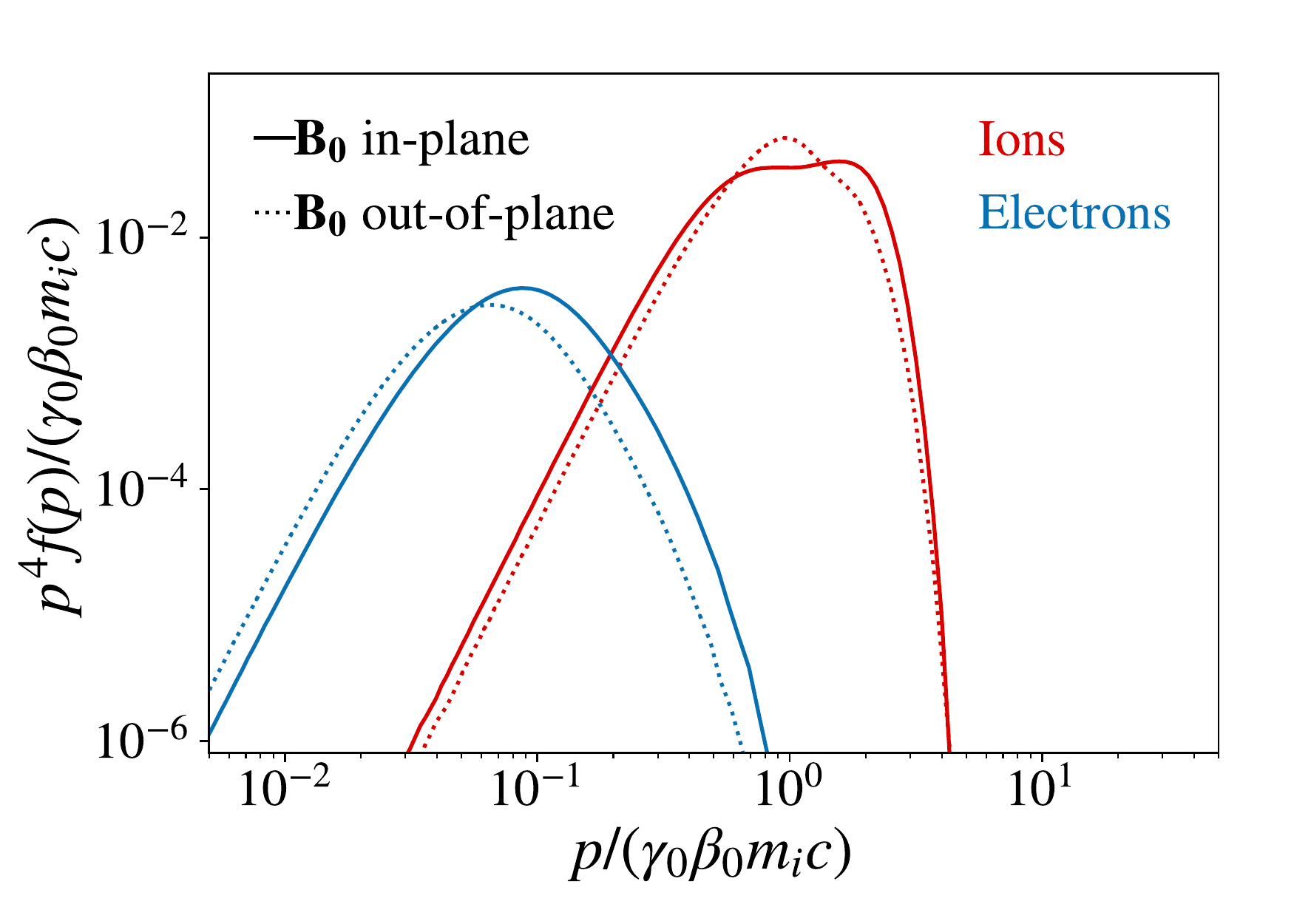}
\caption{Comparison of downstream superluminal spectra with different initial magnetic field orientations with respect to the simulation plane at time $\omega_{pi}t \approx 2600$. Solid lines correspond to superluminal shocks with $\theta_B =55^\circ$ and in-plane $B_0$ field (i.e. $B_{0,\perp}$ along $y$). Dotted lines correspond a superluminal shock with out-of-plane orientation ($B_{0,\perp}$ along $z$). The spectra are extracted 10--80 $c/\omega_{pi}$ downstream of the shock, and are calculated in the downstream rest frame. The upstream is not shown as the spectra are simply Maxwellians with $T=T_0$. Neither configurations show non-thermal acceleration. The in-plane configuration has hotter downstream ions and electrons and a larger SDA bump in the ion spectrum due to its larger compression ratio (see Figure \ref{fig:OFieldsInVOut}).}
\label{fig:OInVOut}
\end{figure}

\begin{figure}
\centering
\includegraphics[width=0.5\textwidth]{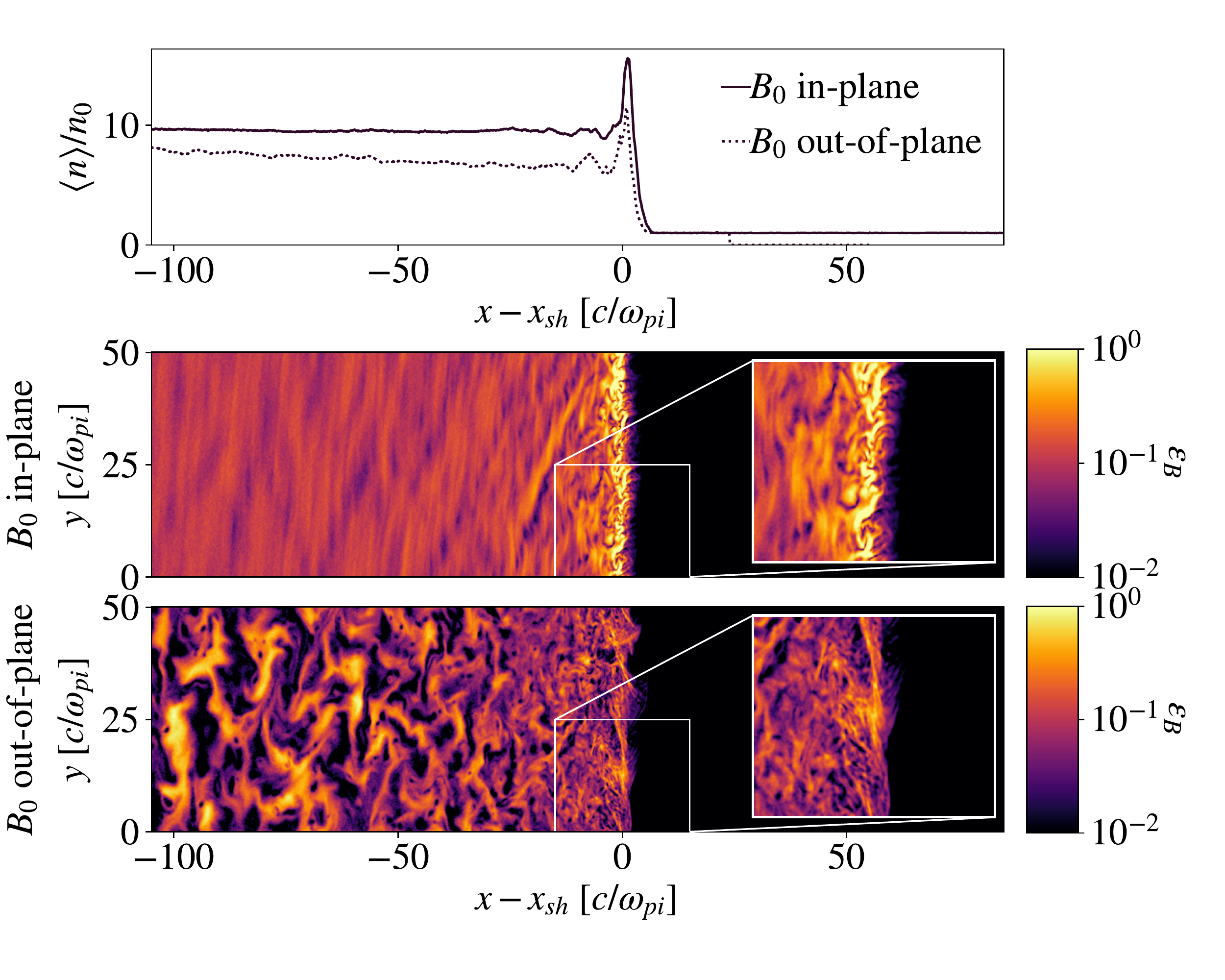}
\caption{Comparison of field quantities close to the shock between superluminal in-plane and out-of-plane magnetic field configuration. As can be seen in the top panel, the two configurations have a different compression ratio, and therefore different shock speeds. The lower panels shows $\epsilon_B$ for the in-plane and out-of-plane configuration. The out-of-plane shock takes longer to thermalize the plasma, showing less compression and a more turbulent downstream. The out-of-plane shock front has structures on the scale of the ion gyroradius. The in-plane shock thermalizes the plasma very quickly and has a higher compression ratio.
}
\label{fig:OFieldsInVOut}
\end{figure}

In this appendix we show that for both the quasi-parallel and superluminal shocks considered in this paper, the orientation of the magnetic field with respect to the simulation plane does not significantly affect our results.

In the main text of the paper, we used an initial magnetic field $B_0$ that was in the $x-y$ plane. There is a growing body of evidence that 2D simulations initialized with $B_{\perp}$ along $z$ as opposed to $y$ may be more efficient at accelerating electrons in high-Mach number, perpendicular shocks \citep[e.g.][]{2012ApJ...755..109M, 2013ApJ...771...54S, 2017ApJ...847...71B}. In the quasi-parallel shocks, the out-of-plane configuration initially reflects more electrons, resulting in an $\epsilon_e$ that is $\sim2$ times larger than the $5\times10^{-4}$ reported in the main paper. However, once the turbulence upstream saturates, the amplified magnetic field is the dominant field at the shock front. Then, the two $B_0$ configurations have approximately the same $\epsilon_e$.

For the superluminal shocks, the field configuration is more important.  The in-plane shock more quickly thermalizes the plasma at the shock front, and therefore has a larger compression ratio. The larger compression ratio results in the ion spectrum having a larger SDA bump, and hotter electrons \citep[see e.g.,][]{2014ApJ...783...91C}. The out-of-plane configuration does not effectively scatter ions in the $z$ direction decreasing the adiabatic index of the shock. The smaller adiabatic index leads to a smaller compression ratio. However, ions do gyrate efficiently in the $x-y$ plane at the shock front, causing more turbulence at the shock front and downstream. In the out-of-plane configuration, the turbulence causes the electrons to be heated far downstream. However, neither configuration is capable of producing upstream magnetic turbulence nor accelerating particles to high energies.
\end{document}